\documentstyle[aas2pp4]{article}
\begin{document}
\title
{General relativistic effects on
neutrino-driven wind from young, hot neutron star and
the r-process nucleosynthesis}

\author{Kaori Otsuki\altaffilmark{1}}
\affil{Division of Theoretical Astrophysics,
National Astronomical Observatory,Mitaka,Tokyo 181-8588}

\author{Hideyuki Tagoshi}
\affil{~~Department of Earth and Space Science, Osaka University,
Toyonaka, Osaka 560-0043}

\author{Toshitaka Kajino\altaffilmark{2,3} and Shin-ya Wanajo}
\affil{Division of Theoretical Astrophysics,
National Astronomical Observatory,Mitaka,Tokyo 181-8588}

\altaffiltext{1}
{Research Center for Nuclear Physics, Osaka University,
Ibaraki,Osaka 560-0043}
\altaffiltext{2}
{Department of Astronomy, University of Tokyo,
Hongo,Tokyo 113-0033}
\altaffiltext{3}
{Department of Astronomical Science,
The Graduate University for Advanced Studies,Mitaka,Tokyo 181-8588}

\begin{abstract}
Neutrino-driven wind from young hot neutron star,
which is formed by supernova explosion,
is the most promising candidate site for
r-process nucleosynthesis.
We study general relativistic effects on this wind in Schwarzschild
geometry in order to look
for suitable conditions for a successful r-process nucleosynthesis.
It is quantitatively discussed that the general relativistic effects
play a significant role in increasing
entropy and decreasing dynamic time scale of the neutrino-driven wind.
Exploring wide parameter region which determines the expansion dynamics
of the wind,
we find interesting physical conditions which lead to successful
r-process nucleosynthesis.
The conditions which we found realize in the neutrino-driven wind with
very short dynamic time scale $\tau_{\rm dyn} \sim 6$ ms and relatively
low entropy $S \sim 140$.
We carry out the $\alpha$-process and r-process nucleosynthesis
calculation on these conditions by the use of our single network code
including over 3000 isotopes, and confirm quantitatively that the second 
and third r-process abundance peaks are produced in the neutrino-driven wind.
\end{abstract}


\section{Introduction}

The r-process is a nucleosynthesis process to produce elements
heavier than iron(\cite{bfh}). They occupy nearly half of the
massive nuclear species, and show typical abundance peaks
around nuclear masses A=80, 130 and 195, whose neutron numbers are
slightly smaller than the magic numbers N=50, 82 and 126, respectively.
This fact suggests that the r-process elements have completely different 
origin from the s-process elements whose abundance peaks are located
just on the neutron magic numbers. The r-process elements are presumed
to be 
produced in an explosive
environment with short time scale and high entropy, where intensive flux of
free neutrons are absorbed by seed elements successively to form the
nuclear reaction flow on extremely unstable nuclei in neutron-rich side.
Recent progress in the studies of nuclear physics of unstable nuclei has 
made it possible to simulate the r-process nucleosynthesis by the use of 
accumulated knowledge on nuclear masses and beta half-lives of several
critical radioactive elements.

The studies of r-process elements make another impact on the cosmic age
problem, that is the age of the Universe to be known from cosmological
constants and the age of the oldest globular cluster conflict with each other. 
A typical r-process element, thorium, has been detected recently in very
metal-deficient stars, providing independent method to
estimate the age of the Milky Way Galaxy(\cite{sn}).
Since thorium has half-life of 14 Gyr, the observed abundance relative
to the other stable elements is used as a chronometer
dating the age of the Galaxy.
To study the origin of the r-process elements is thus important and even
critical in cosmology and astronomy of  Galactic chemical evolution as
well as nuclear physics of unstable nuclei.
Unfortunately, however, astrophysical site of the  r-process
nucleosynthesis has been poorly known, although several candidate sites
are proposed and being investigated theoretically.

Neutrino-driven wind, which is our object to study in this article, is
thought to be one of the most
promising candidates for the r-process nucleosynthesis.
It is generally believed that a neutron star is formed as the remnant of
gravitational core collapse of Type II, Ib or Ic supernovae.
The hot neutron star just born releases most of its energy as neutrinos
during Kelvin-Helmholtz cooling phase, and these neutrino drive
matter outflow from the surface.
This outflow is called neutrino-driven wind.
Many theoretical studies of neutrino-driven wind followed the successful
detection of energetic 
neutrinos from SN1987A, which raised the possibility of finding the
r-process nucleosynthesis in this wind.  

Although there are several numerical simulations of the neutrino-driven
wind, results are very different from one another, depending on models
and methods adopted in literature (\cite{wil,jan1,jan2}).
A benchmark study of numerical simulation by Wilson and his
collaborators (\cite{wil})
can successfully explain the solar system r-process abundances, but the
others (\cite{jan1,jan2}) can not reproduce 
their result. 
Qian and Woosley (1996) tried to work out this discrepancy 
using approximate methods to solve the spherically 
symmetric, steady state flow in the Newtonian framework.

They could not find suitable condition for the r-process
nucleosynthesis, and they suggested in a post-Newtonian calculation that
general relativistic 
effects may improve thermodynamic condition for the r-process nucleosynthesis.
Cardall and Fuller (1997) adopted similar approximate methods in
general relativistic framework and obtained short 
dynamic time scale of the expansion and large
entropy, which is in reasonable agreement with the result in
post-Newtonian approximation adopted by Qian and Woosley (1996).
They did not remark quantitatively, however, what kind of specific
effect among several general relativistic effects is responsible for
this change.

Since the wind blows near the surface of the neutron star, it is needed to
study expansion dynamics of neutrino-driven wind
in general relativity.
The first purpose of this paper is to quantitatively make clear the
effects of general 
relativity by adopting fully general relativistic framework.
Although we assume only spherical steady-state flow of the
neutrino-driven wind, we do not adopt approximate 
methods as in several previous studies. 
We try to extract as general properties as possible of the wind in
manners independent of supernova models so that they are to be compared with
expansion of different object like accretion disk of binary neutron star 
merger (\cite{nsm}) or sub-critical small mass neutron star
(\cite{sumi2}), which is induced by intense neutrino burst.  
The second purpose is to look for suitable conditions for the r-process.
There are key quantities in order to explain the solar system r-process
abundances.
They are the mass outflow rate, $\dot{M}$, the dynamic time scale of the
expansion, $\tau_{\rm{dyn}}$, the entropy, $S$, and the electron fraction,
$Y_{\rm e}$.
The third purpose of this paper is to make clear how these thermodynamic 
and hydrodynamic quantities affect the r-process nucleosynthesis by
carrying out the nucleosynthesis calculation numerically.

In the next section we explain our theoretical models
of neutrino-driven wind.
We introduce basic equations to describe the dynamics of the wind
in the Schwarzschild geometry.
Boundary conditions and adopted parameters for solving these equations
are presented in this section.
Numerical results are shown in section 3, where the effects of
general relativity are studied in detail.
We also investigate the dependence of
the key physical quantities like $\tau_{dyn}$ and $S$ on the neutron
star mass, radius, and neutrino luminosity in order to look for the
conditions of the neutrino-driven wind which is suitable for the
r-process nucleosynthesis. 
Applying the result obtained in section 3, we carry out the
nucleosynthesis calculation in section 4. The purpose of this section is 
to confirm quantitatively that the r-process elements are produced
successfully in the wind having very short dynamic time scale with
relatively low entropy.  
We finally summarize the results of this paper
and present further
discussions and outlook in section 5.

\section{Models of neutrino-driven winds}
\subsection{Basic equations}

Type II or Ib supernova explosion is one of the complex hydrodynamic
process which needs careful theoretical studies of the convection
associated with shock propagation.
The time of our interest, however, is the later phase after the core
bounce, at which the shock has already passed away to reach a radius
about 10000 km and continuous mass outflow is installed from the surface 
of the neutron star.
Recent three dimensional numerical simulation (Hillebrandt 1998) has
indicated that the convection near the shock front does not grow as deep 
as that shown in two-dimensional numerical simulation and the
hydrodynamic conditions behind the shock are more likely similar to
those obtained in one-dimensional numerical simulation.
Since Wilson's numerical simulation of SN1987A in Woosley et al.~(1994)
has shown that the neutrino-driven wind is adequately described by a
steady state flow, we here adopt spherically symmetric and steady state
wind, following the previous studies (\cite{dun,qw,cf}).
According to his numerical simulation, the neutrino luminosity $L_{\nu}$ 
changes slowly from about $10^{52}$ ergs/s to below $10^{51}$
ergs/s during $\sim 10$ s of the Kelvin-Helmholtz cooling phase of the neutron
star.
The properties of the protoneutron star, {\it i.e.} the mass $M$ and
radius $R$, also evolve slowly.
We therefore take these quantities $L_{\nu}$, $M$, and $R$ as input
parameters in order to describe more rapid evolution of the
neutrino-driven wind. 

The basic equations to describe
the spherically symmetric and steady state winds
in Schwarzschild geometry are given by (\cite{st})
\begin{equation}
\dot{M} = 4\pi r^2\rho_{b}u,
\label{eqn:gk1}
\end{equation}
\begin{equation}
u\frac{du}{dr}= \frac{1}{\rho_{tot}+P}\frac{dP}{dr}
\left(1+u^2-\frac{2M}{r}\right)-\frac{M}{r^2},
\label{eqn:gk2}
\end{equation}
\begin{equation}
\dot{q}= u\left( \frac{d\varepsilon}{dr}-\frac{P}{\rho_{b}^{2}}
\frac{d\rho_{b}}{dr}\right),
\label{eqn:gk3}
\end{equation}
where $\dot{M}$ is the mass outflow rate, $r$ is the distance from the
center of the neutron star, $\rho_{b}$ is the baryon mass
density, $u$ is the radial component of the
four velocity, $\rho_{tot}=\rho_{b}+\rho_{b}\varepsilon$ is total energy
density, $\varepsilon$ is the specific internal energy, $P$ is the
pressure, 
$M$ is the mass of the neutron star, and $\dot{q}$ is
the net heating rate due to neutrino interactions with matter.
We use the conventional units that the plank constant $\hbar$, the speed 
of light $c$, the Boltzmann constant $k$, and gravitational constant $G$, 
are taken to be unity. 
Since the neutrino-driven wind blows from the surface of the hot
protoneutron star at high temperature $T \sim 5$ MeV and also the physics of
the wind is mostly determined at $T \gtrsim 0.5$ MeV (\cite{qw}),
the equations of state are approximately written as 
\begin{eqnarray}
P&=&{11\pi^2\over 180}T^4+{\rho_b\over m_N}T,
\label{eqn:eos1}
\\
\varepsilon&=&{11\pi^2\over 60}{T^4\over \rho_b}+
{3\over 2}{T\over m_N},
\label{eqn:eos2}
\end{eqnarray}
where $T$ is the temperature of the system, and $m_N$ is the nucleon rest mass.
We have assumed that the material in the wind consists of photons, relativistic
electrons 
and positrons, and non-relativistic free nucleons.

The heating rate $\dot{q}$ in Eq.~(\ref{eqn:gk3}) 
through the interactions between neutrinos and material
takes the key to understand the dynamics of the neutrino-driven wind.
Following Bethe (1993) and Qian and Woosley (1994), we take account of
the following five neutrino processes;
neutrino and antineutrino absorption by free nucleons,
neutrino and antineutrino scattering by electrons and positrons,
and neutrino-antineutrino annihilation into electron-positron
pair as the heating processes, and
electron and positron capture by free nucleons, and
electron-positron annihilation into neutrino-antineutrino
pair as the cooling processes.
We assume that neutrinos are emitted isotropically from the surface of
the neutron star at the radius $R$, which proves to be a good approximation 
in recent numerical studies of the neutrino transfer (\cite{yamada}).
In this paper, therefore, we make an assumption that the neutrinosphere
radius is equal to the protoneutron star radius $R_{\nu} = R$.
Since the neutrino trajectory is bent in the Schwarzschild geometry,
the material in the wind sees neutrinos within the solid angle
subtended by the neutrinosphere 
which is greater than the solid angle in the Newtonian geometry 
at the same coordinate radius.
The bending effect 
of the neutrino trajectory increases the heating rate 
compared to Newtonian case. We have to
take account of the redshift effect on  the neutrino energy, too, which
tends to decrease the heating rate.

The important heating rate is due to the neutrino and antineutrino
absorption by free nucleons 
\begin{equation}
\nu_e + n \rightarrow p + e^- ,
\label{eqn:mib}
\end{equation}
\begin{equation}
\bar{\nu}_e + p \rightarrow n + e^+ ,
\label{eqn:plb}
\end{equation}
and it is given by
\begin{eqnarray}
\dot{q}_1&\approx&
9.65N_A[(1-Y_e)L_{\nu_e,51}\varepsilon^2_{\nu_e}+Y_eL_{\bar{\nu}_e,51}
\varepsilon^
2_{\bar{\nu}_e}]
\nonumber
\\
&\times&\frac{1-g_1(r)}{R^2_{\nu
    6}}\Phi(r)^6\rm{MeV~ s^{-1}g^{-1}} ,
\label{eqn:q1}
\end{eqnarray}
where the first and second terms in the parenthesis are for the processes
(\ref{eqn:mib}) and (\ref{eqn:plb}), respectively, $\varepsilon_i$ is
the energy in MeV defined by 
$\varepsilon_i=\sqrt{<E_i^3>/<E_i>}$, and $<E_i^n>$
denotes the $n$th energy moment of the neutrino $(i=\nu _e)$ and 
antineutrino $(i=\bar{\nu}_e)$ energy 
distribution, $N_A$ is the Avogadro number,
$Y_e$ is the electron fraction, $L_{i,51}$ is the individual neutrino or
antineutrino luminosity in units of $10^{51}$ ergs/s, and $R_{\nu6}$ is
the neutrinosphere radius in units of $10^6$ cm.
In this equation, $1-g_1(r)$ is the geometrical factor which represents
the effect of bending neutrino trajectory, and $g_1(r)$ is given by
\begin{equation}
g_1(r)=\left(1-\left(\frac{R_{\nu}}{r}\right)^2
\frac{1-2M/r}{1-2M/R_{\nu}}\right)^{1/2},
\end{equation}
where the function $(1-2M/r)/(1-2M/R_{\nu})$ arises due to the
Schwarzschild geometry, and unity should be substituted for this factor
in the Newtonian geometry.  
We also define the redshift factor 
\begin{equation}
\Phi(r)=\sqrt{\frac{1-2M/R_\nu}{1-2M/r}},
\label{eqn:red}
\end{equation}
in the Schwarzschild geometry, which is unity in the Newtonian geometry.
We will discuss the effects of these general relativistic correction
factors in the next section.

The second heating rate due to neutrino and antineutrino scattering by
electrons and positrons plays equally important role.
Neutrinos of all flavors can contribute to the scattering, and the
heating rate is given by
\begin{eqnarray}
\dot{q}_3 &\approx&
2.17N_A\frac{T^4_{MeV}}{\rho_8}
\nonumber
\\
&\times&\left(L_{\nu_e,51}\epsilon_{\nu_e}+L_
{\bar{\nu}_
e,51}\epsilon_{\bar{\nu}_e}+\frac{6}{7}L_{\nu_{\mu},51}
\epsilon_{\nu_{\mu}}
\right)
\nonumber
\\
&\times&
\frac{1-g_1(r)}{R^2_{\nu
    6}}\Phi(r)^5\rm{ MeV~ s^{-1}g^{-1}},
\label{eqn:q3}
\end{eqnarray}
where $\epsilon_i=<E_i^2>/<E_i>$ in MeV
$(i=\nu_e,~~\bar{\nu}_e,~~and~~\nu_{\mu} )$, and we have assumed the same
contribution from $\nu_{\mu}$, $\bar{\nu}_{\mu}$, $\nu_{\tau}$, and
$\bar{\nu}_{\tau}$ fluxes.
We take $\varepsilon_i^2\simeq 1.14\epsilon_i^2$ from the numerical
studies by Qian and Woosley (1996). 

The third heating rate due to neutrino-antineutrino pair annihilation
into electron-positron pair is given by
\begin{eqnarray}
\dot{q}_5 &\approx&
12.0N_A
\nonumber
\\
&\times&
\left(L_{\nu_e,51}L_{\bar{\nu}_e,51}(\epsilon_{\nu_e}+\epsilon_
{\bar{\nu}_e})+\frac{6}{7}L^2_{\nu_{\mu},51}\epsilon_{\nu_{\mu}}\right)
\nonumber
\\
&\times&
\frac{g_2(r)}
{\rho_8 R^4_{\nu 6}}\Phi(r)^9\rm{MeV~ s^{-1}g^{-1}},
\label{eqn:q5}
\end{eqnarray}
where $g_2(r)$ is given by
\begin{equation}
g_2(r)=(1-g_1(r))^4(g_1(r)^2 + 4 g_1(r) + 5).
\label{eqn:geon}
\end{equation}

The cooling rates which we included in the present calculations are for
the inverse reactions of the two heating processes considered in
Eqs.~(\ref{eqn:q1}) and (\ref{eqn:q5}).
The first cooling rate due to electron and positron captures
by free nucleons, which are the inverse
reactions of (\ref{eqn:mib}) and (\ref{eqn:plb}), is given by
\begin{equation}
\dot{q}_2 \approx 2.27N_AT^6_{MeV}\rm{MeV~ s^{-1}g^{-1}}.
\label{eqn:q2}
\end{equation}
The second cooling rate due to electron-positron pair annihilation into
neutrino-antineutrino pair of all flavors, which is the inverse reaction 
of Eq.~(\ref{eqn:q5}), is given by
\begin{equation}
\dot{q}_4 \approx 0.144N_A\frac{T^9_{MeV}}{\rho_8}\rm{ MeV~ s^{-1}g^{-1}}.
\label{eqn:q4}
\end{equation}

Combining the above five heating and cooling rates, we obtain the total
net heating rate $\dot{q}$ 
\begin{equation}
\dot{q}=\dot{q}_1 - \dot{q}_2 + \dot{q}_3 - \dot{q}_4 + \dot{q}_5.
\label{eqn:qtot}
\end{equation}
As we will discuss in the next section, the first three heating and
cooling rates $\dot{q}_1$, $\dot{q}_2$, and $\dot{q}_3$ dominate over
the other two contributions from $\dot{q}_4$ and $\dot{q}_5$. 

\subsection{Boundary conditions and input parameters}

We assume that the wind starts from the surface of the protoneutron star
at the radius $r_i=R$ and the temperature $T_i$.
Near the neutrinosphere and the neutron star surface, both heating
(mostly $\dot{q}_1$) and cooling (mostly $\dot{q}_2$) processes almost
balance with each other due to very efficient neutrino interactions with 
material.
The system is thus in kinetic equilibrium (Barrows and Mazurek 1982) at
high temperature and high density. 
The inner boundary temperature $T_i$ is determined so that
the net heating rate $\dot{q}$ becomes zero at
this radius. 
We have confirmed quantitatively that a small change in $T_i$ does not
influence the calculated thermodynamic and hydrodynamic quantities of the
neutrino-driven wind very much.
We give the density  $\rho(r_i)=10^{10}$ g/cm$^3$ at the inner boundary, 
which is taken from the result of Wilson's numerical simulation in
Woosley et al.~(1994). 

The luminosity of each type of neutrino $L_i
~~(i=\nu_e,~\bar{\nu}_e,~\nu_{\mu},~\bar{\nu}_{\mu},~\nu_{\tau},~\bar{\nu}_{\tau} 
)$ is similar to one another and changes from about $10^{52}$ to
$10^{50}$ ergs/s very slowly during $\sim 10$ s (\cite{wil}). 
We therefore take a common neutrino luminosity $L_{\nu}$ as a constant
input parameter. 
In the heating and cooling rates, however, we use the values of neutrino
energies $\epsilon_{\nu_e}=12~\rm{MeV}$,
$\epsilon_{\bar{\nu_e}}=22~\rm{MeV}$, and
$\epsilon_{\nu}=\epsilon_{\bar{\nu}}=34~\rm{MeV}$ for the other flavors
at $r_i=R$ as in Qian \& Woosley(1996).  
We take the neutron star mass as a constant input parameter ranging $1.2 
M_{\odot} \leq M \leq 2.0 M_{\odot}$, too.

The mass outflow rate $\dot{M}$
determines how much material is ejected by
the neutrino-driven wind.
In Eqs.(\ref{eqn:gk1})-(\ref{eqn:gk3}), $\dot{M}$ is taken to be a 
constant value to be determined by the following outer boundary
condition.
In any delayed explosion models of Type II supernovae ~(\cite{wil,jan1,jan2}),
the shock wave moves away at the radius around 10000 km
above the neutron star surface at times $1 {\rm s} \lesssim t$ after the 
core bounce. 
As we stated in the previous subsection, the neutrino-driven wind is
described by a steady state flow fairly well between the neutron star
surface and the shock.
\ From this observation, a typical temperature at the location of the
shock wave can be used as an outer boundary condition.
We impose the boundary condition only for
subsonic solutions by choosing the value of $\dot{M} < \dot{M}_{\rm
crit}$ so that $T=0.1$MeV at $r\simeq$10,000km, where $\dot{M}_{\rm
crit}$ is the critical value for supersonic solution.
Given $\rho(r_i)$, Eq.(\ref{eqn:gk1}) determines also the initial velocity at
$r=r_i$ for each $\dot{M}$.

We here explore the effects of the assumed boundary condition
and the mass outflow rate $\dot{M}$ on the results of 
calculated quantities of the neutrino-driven winds.
We show in Figs.~\ref{fig1}(a) and ~\ref{fig1}(b) the fluid velocity and
the temperature  
as a function of raduis from the center of neutron star
for various $\dot{M}$, where neutron star mass $M = 1.4~M_{\odot}$ 
and neutrino luminosity $L_{\nu_e} = 10^{51}$ ergs/s are used. 
Figures ~\ref{fig2}(a) and ~\ref{fig2}(b) are the same as those in Figs.
\ref{fig1}(a) and ~\ref{fig1}(b)
for M = 2.0$M_{\odot}$ and $L_{\nu_e} = 10^{52}$ ergs/s. 
Varied $\dot{M}$'s are tabulated in Table 1 with the calculated
entropies and dynamic timescales.
These figures indicate that both velocity and temperature profiles
are very sensitive to the adopted $\dot{M}$ corresponding to
different boundary conditions at r = 10000 km. 
However, the entropies are more or less similar to one another, 
while exhibiting very different dynamic timescales.

Although finding an appropriate boundary condition is not easy,
it is one of preferable manners to match 
the condition obtained in numerical simulations of 
the supernova explosion. 
We studied one of the successful simulations of 20$M_{\odot}$ supernova
explosion assuming $M = 1.4 M_{\odot}$ 
(\cite{wil2}).
Extensive studies of the r-process (\cite{wil}) are based
on his supernova model.  Careful observation tells us that, 
although the neutrino luminosity for each flavor changes 
from $5\times 10^{52}$ ergs/s to $10^{50} $ergs/s,
the temperature lowers progressively to 0.1 MeV around 
r = 10,000 km where the shock front almost stays during $\sim$ 10 s 
after the core bounce at times which we are most interested in.
It is to be noted that for successful r-process (\cite{wil}) 
the temperature has to decrease gradually 
down to around 0.1 MeV at the external region. 
This will be discussed in later sections. 
As displayed in Figs. ~\ref{fig1}(a) and ~\ref{fig1}(b), our calculation 
denoted by "3" meets with this imposed boundary condition.
Although it may not be necessarily clear, we can adopt the same boundary 
condition for different neutron star masses which we study 
in this article, expecting that the physics continuously changes 
and also aiming at comparing the results with one another
which arise from the same boundary condition. 
Even in the case of massive neutron star having M = 2.0$M_{\odot}$,
as displayed in Figs. \ref{fig2}(a) and \ref{fig2}(b), we can still find a
solution
denoted by "1" which satisfies the same outer boundary condition.
Although we fortunately found a solution with reasonable value of 
$\dot{M}$, careful studies of the numerical simulation in the case of 
massive neutron stars are highly desirable in order to 
find better boundary condition.

Let us discuss how our adopted outer boundary condition is 
not unresonable.
We are interested in the times 
$1 {\rm s} \lesssim t$ when the neutrino-driven wind becomes quasi
steady state flow between the neutrinosphere and the shock front. 
Intense flux of neutrinos from the hot proto-neutron star have already 
interacted efficiently with radiation and relativistic electron-positron 
pairs at high temperature. 
Thus we have  $T \sim T_{\nu}$, where $T$ and $T_{\nu}$ are 
respectively the photon and neutrino temperatures.
In this stage, the gain radius $R_g$ (\cite{bw})
at which the neutrino heating and cooling balance with each other
is very close to the neutrinosphere.  Since we make an approximation 
that the neutrinosphere and the neutron star surface is close enough, 
we here assume that the gain radius is also the same, i.e. $R_g = R_{\nu}=R$.
On these conditions we can estimate the mass outflow rate $\dot{M}$
by considering the energy deposition to the gas from the main
processes of neutrino capture on nucleons (\ref{eqn:mib}) and (\ref{eqn:plb}). 

Following the discussion by Woosley et al. (1994), 
the rate of energy deposition in the gas above the neutrinosphere
is given by
\begin{equation}
\dot{E} = (L_{\nu_e} + L_{\bar{\nu}_e}) \times \tau_{\nu},
\label{eqn:edot}
\end{equation}
where $\tau_{\nu}$ is the optical depth for the processes
(\ref{eqn:mib}) and (\ref{eqn:plb}) 
and is given in terms of the opacity $\kappa_{\nu}$ and 
the pressure scale height $L_p$ by, 
\begin{eqnarray}
\tau_{\nu}& =&
\int^{R_g}_{\infty} \kappa_{\nu}\rho_{\rm b}dr
\nonumber
\\ 
&\approx& \kappa_{\nu}(R_g)\rho_{\rm b}(R_g)L_p(R_g)  
\nonumber
\\
&\approx& 0.076 R_7^2 \left(\frac{T_{\nu}}{3.5{\rm MeV}}\right)^6
\left(\frac{1.4M_{\odot}}{M}\right).
\label{eqn:optd}
\end{eqnarray}
Note that $R_g = R$ and $T_{\nu} = T_i$.
In order to obtain this expression, we have already used
an approximate opacity (\cite{brev,ww}) 
$\kappa_{\nu} \approx 6.9\times10^{-18}
(T_{\nu}/3.5{\rm MeV})^2 {\rm cm}^2 {\rm g}^{-1}$
and the pressure scale height in radiation dominated domain 
which is written as,
\begin{eqnarray}
\L_p &\approx& (aT^4)/(GM\rho_{\rm b}/R^2)
\nonumber
\\
&= & 74 {\rm km} \left(\left(\frac{T}{{\rm MeV}}\right)^4R_7^2/\rho_{\rm
b,7}\right)\left(\frac{1.4M_{\odot}}{M} \right),
\nonumber
\\
&&
\label{eqn:scht}
\end{eqnarray}
where the subscripts on $R_7$ and $\rho_{\rm b,7}$ indicate
cgs multipliers in units of $10^7$.
The energy deposition Eq.(\ref{eqn:edot}) is mostly used for
lifting the matter out of the gravitational well of the neutron star.
Thus, inserting Eq.(\ref{eqn:optd}) into Eq.(\ref{eqn:edot})
and using the relation $L_{\nu_e} = L_{\bar{\nu_e}}$
= $(7/4)\pi R^2 \sigma T_{\nu}^4$, 
the mass outflow rate $\dot{M}$ is approximately given by
\begin{eqnarray}
\dot{M} &\approx& \dot{E}/(GM/R) 
\nonumber
\\
&\approx& 0.092\left(\frac{L_{\nu_e} + L_{\bar{\nu_e}}}{10^{53} {\rm ergs}~ {\rm
s}^{-1}}\right)^{5/2}
\left(\frac{1.4M_{\odot}}{M}\right)^2 M_{\odot}{\rm s}^{-1}.
\nonumber
\\
\label{eqn:M}
\end{eqnarray}
Our mass outflow rate $\dot{M}$ obtained from the imposed boundary 
condition of a temperature 0.1 MeV at 10,000 km is in reasonable 
agreement with the estimate using this Eq.~(\ref{eqn:M})
within a factor of five for 
$10^{50} {\rm ergs}/{\rm s} \leq (L_{\nu_e} + L_{\bar{\nu_e}}) 
\leq 10^{52} {\rm ergs}/{\rm s}$.

\subsection{Characteristics of the neutrino-driven wind}

When the material of the wind is on the surface of the neutron star and
neutrinosphere, thermodynamic quantities still reflect the effects of
neutralization and the electron fraction $Y_e$ remains as low as
$\sim 0.1$.
Once the wind leaves surface after the core bounce, electron number
density decreases abruptly and the chemical equilibrium among leptons is 
determined by the balance between the two processes (\ref{eqn:mib}) and
(\ref{eqn:plb}) due to intense neutrino fluxes, shifting $Y_e$ to $\sim 0.5$. 
Interesting phase starts when the temperature falls to $\sim 10^{10}$ K,
for our purpose of studying the physical condition of the
neutrino-driven wind that is suitable for the r-process nucleosynthesis.
At this temperature the material is still in the NSE, and the baryon
numbers are carried by only free protons and neutrons.
The neutron-to-proton number abundance ratio is determined by $Y_e$ for
charge neutrality.

Electron antineutrino has a harder spectrum than electron neutrino, as
evident from their energy moments $\epsilon_{\nu _e}=12$ MeV $< \epsilon 
_{\bar{\nu}_e}=22$ MeV.
Thus, the material is slightly shifted to neutron-rich.
Assuming weak equilibrium, this situation is approximately described by 
\begin{eqnarray}
Y_e &\approx& \frac{\lambda _{\nu_e n}}{\lambda _{\nu_e n} + \lambda
_{\bar{\nu}_e p}} 
\nonumber
\\
&\approx&
\left
  ( 1+\frac{L_{\bar{\nu_e}}}{L_{\nu_e}}\frac{\epsilon_{\bar{\nu_e}}-2
  \delta + 1.2 \delta^2 / \epsilon_{\bar{\nu_e}}}{\epsilon_{\nu_e}+2
  \delta + 1.2 \delta^2 / \epsilon_{\nu_e}}
\right)^{-1},
\nonumber
\\
&&
\label{eqn:ye}
\end{eqnarray}
where $\lambda_{\nu _e n}$ and $\lambda_{\bar{\nu}_e p}$ are the
reaction rates for the processes (\ref{eqn:mib}) and (\ref{eqn:plb}),
respectively, and  $\delta$ is the neutron-proton mass difference
(\cite{qw}).
In our parameter set of the neutron star mass $M=1.4M_{\odot}$ and
radius $R=10$ km, for example, $Y_e$ varies from $Y_e(r=R)=0.43$ to
$Y_e(r=10000 {\rm km})=0.46$ very slowly due to the redshift factor
(\ref{eqn:red}) because of $\epsilon \propto \Phi$.
As this change is small and the calculated result of hydrodynamic
quantities are insensitive to $Y_e$, we set $Y_e=0.5$ for numerical
simplicity.

One of the most important hydrodynamic quantity, that characterizes the
expansion dynamics of the neutrino-driven wind, is the dynamic time
scale $\tau_{\rm dyn}$ which is the duration time of the
$\alpha$-process.
When the temperature falls below $10^{10}$ K, the NSE favors a
composition of alpha-particles and neutrons.
As the temperature drops further below about $5 \times 10^{9}$ K ($T
\approx 0.5$ MeV), the system falls out of the NSE and the
$\alpha$-process starts accumulating some amount of seed elements until
the charged particle reactions freeze out at $T \approx 0.5/e$ MeV
$\approx 0.2$ MeV.
Introducing a time variable of the wind moving away from the distance
$r_i$ to outer distance $r_f$
\begin{equation}
\tau=
\int^{r_f}_{r_i}\frac{dr}{u},
\label{eqn:time}
\end{equation}
and setting $r_i=r(T=0.5 {\rm MeV})$ and $r_f=r(T=0.5/e~~ {\rm MeV})$, we
can define the dynamic time scale $\tau_{\rm dyn}$ by
\begin{equation}
\tau_{dyn}\equiv
\int^{\rm{T=0.5/e~~MeV}}_{\rm{T=0.5 MeV}}\frac{dr}{u}.
\label{eqn:tau}
\end{equation}

The second important hydrodynamic quantity, that affects strongly the
r-process nucleosynthesis which occurs at later times when the
temperature cools below $0.2$ MeV, is the entropy per baryon, defined by
\begin{equation}
S= \int ^r _R \frac{m_N \dot{q}}{u T}dr,
\label{eqn:S}
\end{equation}
where $\dot{q}$ is the total net heating rate (\ref{eqn:qtot}).
As $S \propto T^3/\rho_b$ assuming the radiation dominance, high entropy 
and high temperature characterizes a system with many photons and low
baryon number density.
Since high entropy favors also a large fraction of free nucleons in the limit
of the NSE, it is expected to be an ideal condition for making high
neutron-to-seed abundance ratio.
Therefore, the high entropy at the beginning of the $\alpha$-process is
presumed to be desirable for successful r-process.

\section{Numerical results}
\subsection{Effects of relativistic gravity to entropy}

The purpose of this section is to discuss both similarities and
differences of the neutrino-driven wind between the relativistic
treatment and the Newtonian treatment.
In Fig.~\ref{fig3}, we show typical numerical results of radial velocity 
$u$, temperature $T$, and baryon mass density $\rho_{b}$ of the wind
for the neutron star mass $M=1.4 M_{\odot}$, radius
$R=10$km, and the 
neutrino luminosity $L_{\nu }=10^{51}$ ergs/s.
The radial dependence of these quantities is displayed by solid and dashed
curves for Schwarzschild and Newtonian cases, respectively, in this figure.
Using these results and Eq.(\ref{eqn:S}), we can calculate $S$ in each ejecta.
Figure \ref{fig4} shows the calculated profile of the entropy $S$ for
the two cases. 
Although both entropies describe rapid increase just above the surface
of the neutron star $10$ km $\leq r \leq 15$ km, the asymptotic value
in general relativistic wind is nearly $40$ \% larger than that in
Newtonian wind.

The similar behavior of rapid increase in both winds is due to
efficient neutrino heating near the surface of the neutron star.
We show the radial dependence of the heating and cooling rates by
neutrinos in Figs.~\ref{fig5}(a)-(c).
Figure ~\ref{fig5}(a) shows the total net heating rate defined by
Eq.~(\ref{eqn:qtot}), and Figs.~\ref{fig5}(b) and (c) display the
decompositions into contribution from each heating(solid) or
cooling(dashed) rate in Schwarzschild and Newtonian cases,
respectively.
The common characteristic in both cases is that net heating rate
$\dot{q}$ has a peak around 
$r \approx 12$ km, which makes a rapid increase in $S$ near the surface
of the neutron star for the following reason.
The integrand of the entropy $S$ in Eq.~(\ref{eqn:S}) consists of
the heating rate and the inverse of fluid velocity times temperature.
The fluid velocity increases more rapidly than the slower decrease in
the temperature, as shown in Fig.~\ref{fig3}, after the wind lifts off
the surface of the neutron star.

Let us carefully discuss the reason why the general relativistic wind
results in $40$ \% larger entropy than the Newtonian wind in the
asymptotic region.
This fact has been suggested in the previous papers of Qian \&
Woosley~(1996) and Cardall \& Fuller (1997).
Unfortunately, however, the reason of this difference was not clearly
appreciated to the specific effect quantitatively among several
possible sources.

We first consider the redshift effect and the bending effect of the
neutrino trajectory.
The redshift effect plays a role in decreasing the mean neutrino energy
$\epsilon_{\nu}$ ejected from the neutrinosphere, and in practice
$\epsilon_{\nu}$ is proportional to the redshift factor $\Phi (r)$ which
is defined 
by Eq.~(\ref{eqn:red}).
Since neutrino luminosity is proportional to $\Phi^4$ and
the heating rate $\dot{q}_1, \dot{q}_3$, and $\dot{q}_5$ depend on these
quantities in different manners, each heating rate has different
$\Phi$-dependence as 
$\dot{q}_1 \propto L_{\nu}\epsilon_{\nu}^2 \propto \Phi^6$, 
$\dot{q}_3 \propto L_{\nu}\epsilon_{\nu} \propto \Phi^5$,
and $\dot{q}_5 \propto L_{\nu}^2\epsilon_{\nu} \propto \Phi^9$,
as shown in Eqs. (\ref{eqn:q1}),(\ref{eqn:q3}), and (\ref{eqn:q5}).
Cooling rates $\dot{q}_2$ and $\dot{q}_4$ do not depend on $\Phi(r)$.
The bending effect of the neutrino trajectory is included in the
geometrical factors $g_1(r)$ and $g_2(r)$ in these equations. 
Although numerical calculations were carried out by including all
five heating and cooling processes, 
as $\dot{q}_1$, $\dot{q}_2$, and $\dot{q}_3$ predominate the total
net heating rate  
$\dot{q}$, we here discuss only these 
three processes in the following discussions for simplicity.

In Newtonian analysis, the redshift factor $\Phi(r)$ is unity and the
geometrical factor is given by 
\[g_{1N}(r)=\sqrt{1-\left(\frac{R_{\nu}}{r}\right)^2}.\]
This geometrical factor $g_1(r)$ and the redshift factor appear in the
form of $(1-g_1(r)) \Phi(r)^m$ in the heating rate $\dot{q}_1 ~~(m=5)$
and $\dot{q}_3~~(m=6)$. As for the first factor $(1-g_1(r))$, the
following inequality relation holds between the Schwarzschild and
Newtonian cases, for $R_{\nu} \leq r$;
\[\left( 1-g_1(r)\right) > \left( 1-g_{1N}(r)\right). \]
However, $\Phi (r)$ is a monotonously decreasing function of $r$, the
combined factor 
$(1-g_1(r)) \Phi(r)^m/(1-g_{1N}(r))$ increases from unity and has a
local maximum around $r \sim R_{\nu}+0.2$ km. 
Its departure from unity is at most 3 \% .
Beyond this radius the function starts decreasing rapidly because of the 
redshift effect $\Phi(r)^m$, and it becomes as low as $\sim 0.6$ at $r
\sim 30$ km. In this region,
the net heating rate in the relativistic wind is smaller than
that in the Newtonian wind if the temperature and density are the same.
However, the difference in this region does not influence 
the dynamics of the wind very much.
It is almost determined in
the inner region $R_{\nu} \leq r \lesssim 15$ km where one finds
efficient neutrino heating and small difference between $(1-g_1(r))
\Phi^m(r)$ and $(1-g_{1N}(r))$.

By performing
general relativistic calculation and neglecting
these two relativistic effects, {\it i.e.} the redshift effect and the
bending effect of the neutrino trajectory,
we find that it produces only a small change in entropy
by  $\Delta S \sim 3$.
Thus it does not seem to be the major source
of the increase in the entropy.

Let us consider another source of general relativistic effects which are 
included in the solution of a set of the basic equations
(\ref{eqn:gk1})-(\ref{eqn:gk3}).
Since the entropy depends on three hydrodynamic quantities $\dot{q}(r)$, 
$u(r)$, and $T(r)$ (see Eq.(\ref{eqn:S})), we should discuss each quantity.
The neutrino-heating rate, $\dot{q}(r)$, depends on the temperature
$T(r)$ and density $\rho_{\rm b}(r)$ in addition to the redshift factor
and the geometrical factor of the bending neutrino trajectory.
Therefore, we study first the detailed behavior of $T(r)$, $u(r)$, and
$\rho_{\rm b}(r)$, and then try to look for the reason why the general
relativistic effects increase the entropy.
We assume that the pressure and internal energy per baryon are
approximately described by the radiation and relativistic electrons and
positrons in order to make clear the following discussions.
This is a good approximation for the neutrino-driven wind. 
The equations of state are given by
\begin{equation}
P \approx \frac{11 \pi^2}{180} T^4,
\label{eqn:apn1}
\end{equation}
\begin{equation}
\epsilon \approx \frac{11 \pi^2}{60}\frac{T^4}{\rho_b}.
\label{eqn:apn2}
\end{equation}
By using another approximation
\begin{equation}
u^2 \ll \frac{4P}{3\rho_b},
\label{eqn:apn3}
\end{equation}
which is satisfied in the region of interest, we find 
\begin{eqnarray}
\frac{1}{T} \frac{dT}{dr} &\approx &\frac{1}{1+u^2-\frac{2M}{r}}
\frac{\rho_{\rm b}+P}{4P}
\nonumber
\\
&\times&
\left(-\frac{M}{r^2}
+ \frac{2u^2}{r} - \frac{45}{11 \pi^2}\frac{u \rho_b}{T^4}\dot{q}\right),
\nonumber
\\
&&
\label{eqn:temg}
\end{eqnarray}
in Schwarzschild case.
The basic equations of the spherically symmetric and steady state wind in
Newtonian case are given by
\begin{equation}
\dot{M}=4 \pi r^2 \rho_b v,
\label{eqn:nk1}
\end{equation}
\begin{equation}
v \frac{dv}{d r} =- \frac{1}{\rho_b} \frac{dP}{dr}- \frac{M}{r^2},
\label{eqn:nk3}
\end{equation}
\begin{equation}
\dot{q}=v\left(\frac{d \epsilon}{d r} - \frac{P}{\rho_b^2}\frac{d
    \rho_b}{d r}\right), 
\label{eqn:nk2}
\end{equation}
where $v$ is the fluid velocity. The equations of state are
given by Eqs. (\ref{eqn:eos1}) and (\ref{eqn:eos2}) the same as in
Schwarzschild case.
Repeating the same mathematical technique in
Eqs.~(\ref{eqn:nk1})-(\ref{eqn:nk2}) instead of
Eqs.~(\ref{eqn:gk1})-(\ref{eqn:gk3}) and taking the same approximations
as (\ref{eqn:apn1})-(\ref{eqn:apn3}), we find the equation corresponding to
Eq. (\ref{eqn:temg}), in Newtonian case, as
\begin{equation}
\frac{1}{T} \frac{dT}{dr} \approx \frac{\rho_b}{4P}
\left(-\frac{M}{r^2} + \frac{2v^2}{r} - \frac{45}{11 \pi^2}\frac{v
  \rho_b}{T^4}\dot{q}\right).
\label{eqn:temn}
\end{equation}
Note that the logarithmic derivative of the temperature, $d \ln
T/dr=T^{-1}dT/dr$, has always a negative value, and
  the temperature is a monotonously decreasing function of $r$. 
There are two differences between Eqs.~(\ref{eqn:temg}) and (\ref{eqn:temn}).
The first prefactor $1/(1+u^2-2M/r)$ in the r.h.s. of
  Eq.~(\ref{eqn:temg}) is larger than unity. 
This causes more rapid decrease of $T(r)$ in relativistic case than in
  Newtonian case at small radii within $r \sim 20$ km, as shown in
  Fig.~\ref{fig3}, where our approximations are satisfied.
The second prefactor $(\rho_{\rm b}+P)/4P$ in the r.h.s. of
  Eq.~(\ref{eqn:temg}) is larger than the prefactor $\rho_{\rm b}/4P$ in 
  the r.h.s. of Eq.~(\ref{eqn:temn}), {\it i.e.} $(\rho_{\rm b}+P)/4P >
  \rho_{\rm b}/4P$, which also makes the difference caused by the first
  prefactor even larger.

Applying the similar mathematical transformations to the velocity,
we obtain the following approximations,
\begin{equation}
\frac{1}{u}\frac{du}{dr} \approx \frac{3}{1+u^2-\frac{2M}{r}}
\frac{(\rho_b+4P)}{4P}
\frac{M}{r^2}-\frac{2}{3r}+\frac{\rho_b}{4 u P}\dot{q}
\label{eqn:gv}
\end{equation}
in Schwarzschild case, and
\begin{equation}
\frac{1}{v}\frac{dv}{dr}\approx \frac{3 \rho_b}{4P}\frac{M}{r^2}-
\frac{2}{3r}+\frac{ \rho_b}{4 v P}\dot{q}
\label{eqn:nv}
\end{equation}
in Newtonian case. In these two equations, the first leading term in the 
r.h.s. makes the major contribution. Since exactly the same prefactors
$1/(1+u^2-2M/r)$ and $(\rho_{\rm b}+4P)/4P$ appear in Schwarzschild
case, the same logic as in the logarithmic derivative of the temperature
is applied to the velocity.
Note, however, that slightly different initial velocities at the surface 
of the neutron star make this difference unclear in Fig.~{\ref{fig3}}.
The relativistic Schwarzschild wind starts from $u(10~{\rm km}) \approx 
8.1 \times 10^4$ cm/s, while the Newtonian wind starts from $v(10~{\rm km})
\approx 2.0 \times 10^5$ cm/s.
Both winds reach almost the same velocity around $r \sim 20$ km or beyond.

The baryon number conservation leads to the logarithmic derivative of
the baryon density
\begin{equation}
\frac{1}{\rho_b}\frac{d\rho_b}{dr}=-\frac{1}{u}\frac{du}{dr}
-\frac{2}{r},
\label{eqn:rho}
\end{equation}
where $u$ is the radial component of the four-velocity in Schwarzschild
case.
The fluid velocity $v$ should read for $u$ in Newtonian case.
Inserting Eq.~(\ref{eqn:gv}) or Eq.~(\ref{eqn:nv}) to the first leading 
term of the r.h.s. of this equation, we can predict the behavior of
$\rho_b$ as a function of $r$ in both Schwarzschild and Newtonian cases
as shown in Fig.~\ref{fig3}.

Incorporating these findings concerning $T(r)$, and $u(r)$ into  
the definition of entropy Eq. (\ref{eqn:S}),
we can now discuss why the relativistic Schwarzschild
wind makes larger entropy than the Newtonian wind.
We have already discussed previously in the second paragraph of this section 
that the fluid velocity increases more rapidly in Schwarzschild case.
Since integrand of the entropy $S$ is inversely proportional 
to fluid velocity times temperature, this fact enlarges the difference 
due to $\dot{q}$ at smaller radii (see Fig.~\ref{fig4}(a)).
In addition, as we found, temperature in Schwarzschild geometry is
smaller than the temperature in Newtonian geometry.  For these reasons,
the entropy in the relativistic Schwarzschild wind becomes larger than
the entropy of the Newtonian wind. 

Let us confirm the present results quantitatively in a different manner.
The entropy per baryon for relativistic particles with
zero chemical potential is given by
\begin{equation}
S={11\pi^2\over 45}{T^3\over \rho_b/m_N}.
\label{eqn:sapp}
\end{equation}
Here, we take a common temperature $T=0.5$ MeV to each other in
Schwarzschild and Newtonian cases.
This is the typical temperature at the beginning of the $\alpha$-process,
and both electrons and positrons are still relativistic at this
temperature.
We read off the radii at which 
the temperature becomes 0.5MeV in Fig.~\ref{fig3}.
They are 43km and
55km in Schwarzschild and Newtonian cases, respectively.
We can again read off the baryon mass densities at these radii in this 
figure, that are $\rho_{\rm b}=5.5 \times 10^5$ g/cm$^3$ at $r=43$ km in 
relativistic Schwarzschild wind and $\rho_{\rm b}=7.8 \times 10^5$
g/cm$^3$ at $r=55$ km in Newtonian wind.
Taking the inverse ratio of these $\rho_{\rm b}$ values with approximate 
relation $(\ref{eqn:sapp})$, we find that the entropy in Schwarzschild
case is 40 \% larger than that in Newtonian case.
This is quantitatively in good agreement with the result of numerical
calculation shown in Fig.~\ref{fig4}. 

Let us shortly remark on the dynamic time scale $\tau_{\rm dyn}$.
Although higher entropy is favorable for making enough neutrons in the
neutrino-driven wind, shorter dynamic time scale also is in favor of the 
r-process.
This is because the neutron-to-seed abundance ratio, which is one of the 
critical parameters for successful r-process, becomes larger in the wind 
with shorter $\tau_{\rm dyn}$, which is to be discussed in the next
section.
It is therefore worth while discussing the general relativistic effect
on $\tau_{\rm dyn}$ here.
The argument is very transparent by using Eqs.~(\ref{eqn:temg}) and
(\ref{eqn:temn}) and Fig.~\ref{fig3}.
Since the dynamic time scale $\tau_{\rm dyn}$ is defined as the duration 
of $\alpha$-process in which the temperature of the wind cools from
$T=0.5$ MeV to $T=0.5/e \approx 0.2$ MeV, faster cooling is likely to
result in shorter $\tau_{\rm dyn}$.
Let us demonstrate it numerically.
For the reasons discussed below two Eqs.~(\ref{eqn:temg}) and
(\ref{eqn:temn}), the relativistic fluid describes more rapid decrease
in temperature than the Newtonian fluid as a function of distance $r$.
In fact, the distance corresponding to $T=0.5-0.2$ MeV are $r=43-192$ km 
in Schwarzschild case, and $r=55-250$ km in Newtonian case.
Figure \ref{fig3} tells us that both fluids have almost the same
velocities at these distances, which gives shorter $\tau_{\rm dyn}$ for
the Schwarzschild case than the Newtonian case.
The calculated dynamic time scales are $\tau_{\rm dyn}=0.164$ s for the
former and $\tau_{\rm dyn}=0.213$ s for the latter.

     Before closing this subsection, let us briefly discuss
how the system makes a complicated response to the change in
$T(r)$, $u(r)$ and $\rho_{\rm b}(r)$.
When the temperature decreases rapidly at $10~{\rm km}\leq r \lesssim
20$ km, the major cooling process of the $e^+e^-$ capture by free nucleons, 
$\dot{q}_2$, is suppressed because this cooling rate has rather 
strong temperature dependence, $\dot{q}_2 \propto T^6$. 
In Schwarzschild geometry this suppression partially offsets the
decrease in $\dot{q}_1$ due to the neutrino redshift effect,
though being independent of temperature of the wind.
Another heating source $\dot{q}_3$ due to neutrino-electron scattering
also plays a role in the change of entropy.  
Since $\dot{q}_3$ depends on the baryon density
as well as temperature $\dot{q}_3 \propto T^4/\rho_{\rm b}$,
if the system has a correlated response to decrease 
$\rho_{\rm b}$ strongly with decreasing temperature,
then this might eventually work for the partial increase in entropy.
However, in reality, actual response arises from more complicated 
machanism because $\dot{q}_i$'s
should depend on the solution of dynamic equations (1)-(3) self-consistently 
on adopted proper boundary conditions and input parameters
through the relation
$\dot{q}_1 \propto L_{\nu}\epsilon_{\nu}^2$, 
$\dot{q}_2 \propto T^6$, 
$\dot{q}_3 \propto T^4/\rho_{\rm b} L_{\nu}\epsilon_{\nu}$,
$\dot{q}_4 \propto T^9/\rho_{\rm b}$, and
$\dot{q}_5 \propto \rho_{\rm b}^{-1} L_{\nu}^2\epsilon_{\nu}$.
The neutrino-driven wind is a highly non-linear system.

\subsection{Parameter dependence}

Most of the previous studies of the neutrino-driven wind have been
concentrated on SN1987A, and the parameter set in the theoretical
calculations was almost exclusive.
We here expand our parameter region of the neutron star mass $M$, radius 
$R$, and neutrino luminosity $L_{\nu}$, and investigate widely the
dependence of key quantities, $\tau_{\rm dyn}$ and $S$, 
on these three parameters.
Since the neutron star mass $M$ and radius $R$ are mostly contained
through the form $M/R$ in the basic equations of the system, we only
look at the dependence on $M$ and $L_{\nu}$.

Figures ~\ref{fig6}(a) and \ref{fig6}(b) show the calculated $\tau_{\rm
dyn}$ and $S$ at the beginning of the $\alpha$-process at $T=0.5$ MeV
for various neutron star masses $1.2 M_{\odot} \leq M \leq 2.0
M_{\odot}$.
Closed circles, connected by thick solid line, and open triangles,
connected by thin solid line, are those for the Schwarzschild and
Newtonian cases. In Fig.~\ref{fig6}(a), we plot also two broken lines in 
Newtonian case from the paper~(Qian and Woosley~~1996) which adopted
\begin{equation}
\tau_{\rm dyn} ({\rm QW})= \left.\frac{r}{\upsilon} \right|
_{0.5{\rm MeV}},
\label{eqn:qwt}
\end{equation}
in two limits of the radiation dominance (upper) and the dominance of
non-relativistic nucleon (lower).
In either limit, this $\tau_{\rm dyn}$(QW) is an increasing
function of the neutron star mass and this feature is in reasonable
agreement with our exact solution Eq.(\ref{eqn:tau}).
However, absolute value of (\ref{eqn:qwt})is about half that of the
exact solution in the Newtonian case.

Remarkable difference between Schwarzschild and Newtonian cases is
an opposite response of $\tau_{\rm dyn}$ to the neutron star mass
(Fig.~\ref{fig6}(a)).
General relativistic effects make the dynamic time scale even smaller
with increasing neutron star mass.
We have already discussed the reason why $\tau_{\rm dyn}$ in
Schwarzschild case is smaller than that in Newtonian case by comparing
Eqs.~(\ref{eqn:temg}) and (\ref{eqn:temn}) from each other.
We understand the decrease of $\tau_{\rm dyn}$ as a consequence from the 
fact that the general relativistic effects, which arise from the two
prefactors in the r.h.s. of Eq.(\ref{eqn:temg}), are enlarged by stronger
gravitational force $M/r^2$ with larger $M$.
Similar analysis on the role of the gravitational force is applied to
the discussion of entropy and Eqs.~(\ref{eqn:temg}), (\ref{eqn:gv}), and 
(\ref{eqn:rho}).
Figure~\ref{fig6}(b) displays that the entropy per baryon in
Schwarzschild case makes stronger mass dependence than in Newtonian case.

It is to be noted again that the above features of the mass dependence
are equivalent to those obtained by the change in the neutron star
radius.
Since the radius of protoneutron star shrinks with time in cooling
process, it may work for increasing the entropy and decreasing the
dynamic time scale.

Figures~\ref{fig7}(a) and \ref{fig7}(b) show the dependence of our
calculated $\tau_{\rm dyn}$ and $S$ on the neutrino luminosity ranging
$10^{50} {\rm ergs/s} \leq L_{\nu} \leq 10^{52} {\rm ergs/s}$.
Differing from the mass dependence, both quantities are decreasing
function of $L_{\nu}$ as far as $L_{\nu} \leq 10^{52}$ ergs/s.
This tendency, except for the absolute values, is in reasonable
agreement with approximate estimates~(Qian and Woosley 1996) shown by
broken lines.
This is because larger luminosity makes the mass outflow rate $\dot{M}$
higher through more efficient neutrino heating, which causes bigger
increase in the fluid velocity in addition to moderate increase in
baryon density.
Having these changes in hydrodynamic quantities with the definition of
$\tau_{\rm dyn}$, Eq.(\ref{eqn:tau}), and the definition of $S$,
Eq.(\ref{eqn:S}), we 
understand that both quantities decrease with
increasing neutrino luminosity.

However, if the luminosity becomes larger than $10^{52}$ ergs/s, the
temperature does not decrease as low as 0.1 MeV before the distance
reaches 10000 km because of the effect of too strong neutrino heating.
The dynamic time scale $\tau_{\rm dyn}$ is of order $\sim 10$ s.
In such a very slow expansion of the neutrino-driven wind,
$\alpha$-process goes on and leads to uninteresting
r-process nucleosynthesis.

To summarize this section, we find it difficult to obtain very large
entropy $\sim 400$ for reasonably short dynamic time scale $\tau_{\rm
dyn} \lesssim 0.1$ s, as reported by Woosley et al. (1994), by changing
the neutron star mass $M$ and neutrino luminosity $L_{\nu}$.
However, there are still significant differences 
between our calculated result of $\tau_{\rm dyn}$ and $S$, which are
shown by thick solid lines in Figs.~\ref{fig6}(a)-\ref{fig7}(b), and
those of Qian and Woosley (1996), which are shown by broken lines,
in the mass dependence
of the entropy and the opposite behavior in $\tau_{\rm dyn}$.
We will see in the subsequent sections that these differences are
important to look for successful condition of the r-process.   

\subsection{Implication in nucleosynthesis}
Having known the detailed behavior of dynamic time scale $\tau_{\rm dyn}$
and entropy per baryon $S$ as a function of neutron star mass $M$,
radius $R$, and 
neutrino luminosity $L_{\nu}$, we are forced to discuss their
implication in the r-process nucleosynthesis.
We have already shown the calculated results of $\tau_{\rm dyn}$ and $S$ 
for limited sets of two independent parameters $M$ and $L_{\nu}$ in
Figs.~\ref{fig6}(a)-\ref{fig7}(b). 
We here expand the parameter space in order to include a number of
$(M, L_{\nu})$-grids in their reasonable range $1.2 M_{\odot} \leq M
\leq 2.0 M_{\odot}$ and $10^{50} {\rm ergs/s} \leq L_{\nu} \leq 10^{52}$ 
ergs/s.

Figure~\ref{fig8} displays the calculated results in the $\tau_{dyn}-S$ plane.
Shown also are two zones for which the r-process nucleosynthesis might
occur so that the second abundance peak around $A=130$ and the third
abundance peak around $A=195$ emerge from a theoretical calculation as
suggested by Hoffman et al. (1997).
Their condition for the element with mass number $A$ to be produced in
an explosive r-process nucleosynthesis, for $Y_e > \langle Z \rangle /
\langle A \rangle$, is given by  
\begin{equation}
S\approx Y_{e,i} 
\left\{ \frac{8 \times 10^7 (\langle A \rangle -2
\langle Z \rangle)}{\ln[(1-2 \langle Z \rangle /A)/(1 -
\langle A \rangle /A)]} 
\left(\frac{\tau_{dyn}}{sec}\right)\right\}^{1/3},
\label{eqn:hofcon}
\end{equation}
where $\langle A \rangle$ is mean mass number and $\langle Z \rangle$ is
mean proton number of the seed nuclei at the end of the $\alpha$-process.
Following numerical survey of seed abundance of Hoffman et al. (1997),
we choose $\langle A \rangle=90$ and $\langle Z \rangle=34$ in Fig.~\ref{fig8}.
\ From this figure, we find that dynamic time scale as short as $\tau_{\rm 
  dyn} \approx 6$ ms  with 
$M=$2.0$M_{\odot}$ and $L_{\nu}=10^{52}$ergs/s
is the best case among those studied in the present paper in order to
produce the r-process elements,
although the entropy $S$ is rather small 140.

Let us remark shortly on this useful equation.
Equation~(\ref{eqn:hofcon}) tells us that the r-process element with mass
number $A$ is efficiently produced from seed elements with $\langle A
\rangle$ and $\langle Z \rangle$ on a given physical condition
$\tau_{dyn}$, $S$, and $Y_e$ at the onset of r-process
nucleosynthesis at $T_9 \approx 2.5$.
In order to derive Eq.~(\ref{eqn:hofcon}), Hoffman et al. (1997)
assumed that the $\alpha + \alpha + n
\rightarrow ^9{\rm Be} + \gamma$ reaction is in
equilibrium, because of its low Q-value, during the
$\alpha$-process at $T \approx 0.5-0.2$ MeV and that the $^9{\rm Be} +
\alpha \rightarrow ^{12}{\rm C} + \gamma$ reaction triggers burning of
alpha-particles to accumulate seed elements.
The NSE holds true if the nuclear interaction time scale for  $\alpha +
\alpha + n \rightarrow ^9{\rm Be} + \gamma$ is much shorter than the
expansion time scale.
We found in the present calculation that it is not always the case in
neutrino-driven winds with short dynamic time scale, 
for $L_{\nu} \approx 5 \times 10^{51}-10^{52}$ ergs/s, which is to be discussed
more quantitatively in the next section. 
Keeping this in mind, we think that Eq.~(\ref{eqn:hofcon}) is still a
useful formula in order to search for suitable physical condition for
the r-process without performing numerical nucleosynthesis calculation. 

One might wonder if the dynamic time scale $\tau_{\rm dyn} \sim 6$ ms 
is too short for the wind to be heated by neutrinos.
Careful comparison between proper expansion time and specific
collision time for the neutrino heating is needed in order to
answer this question.
Note that $\tau_{\rm dyn}$ was defined as the duration of the
$\alpha$-process so that the temperature of the expanding wind decreases 
from $T=0.5$ MeV to $0.5/e \approx 0.2$ MeV, which correspond to outer
atmosphere of the neutron star. These radii are $r(T=0.5 ~{\rm MeV})=52$
km and $r(T=0.5/e ~{\rm MeV})=101$ km for the wind with $(L_{\nu},
M)=(10^{52} {\rm ergs/s}, 2.0 M_{\odot})$, and $r(T=0.5 ~{\rm MeV})= 43$
km and  $r(T=0.5/e ~{\rm MeV})=192$ km for the wind with $(L_{\nu},
M)=(10^{51} {\rm erg/s}, 1.4 M_{\odot})$.
We found in Figs.~\ref{fig5}(a)-(c) that the neutrinos transfer their
kinetic energy to the wind most effectively just above the neutron star
surface at $10 {\rm km} \leq r < 20 {\rm km}$. 
Therefore, as for the heating problem, one should refer the duration of
time for the wind to reach the radius where temperature is $T \approx
0.5$ MeV rather than $\tau_{\rm dyn}$. 
We can estimate this expansion time $\tau_{\rm heat}$ by 
setting $r_i = R =10$ km and $r_f = r(T=0.5 {\rm MeV})$ in
Eq.~(\ref{eqn:time}): 
$\tau_{\rm heat} = 0.017$ s and $0.28$ s for the winds
with $(L_{\nu},M)=(10^{52} {\rm ergs/s}, 2.0 M_{\odot})$ and $(10^{51}
{\rm ergs/s}, 1.4 M_{\odot})$, respectively.
We note, for completeness, $r(T=0.5 {\rm MeV})=52$ km or 43 km for each 
case.

These proper expansion time scales, $\tau_{\rm heat}$, are to be compared 
with the specific collision time $\tau_{\nu}$ for the neutrino-nucleus
interactions in order to discuss the efficiency of the neutrino heating.
The collision time $\tau_{\nu}$ is expressed (\cite{qhl}) as
\begin{eqnarray}
\tau_{\nu} &\approx& 0.201 \times L_{\nu ,51}^{-1} 
\nonumber
\\
&\times&
\left
  ( \frac{\epsilon _{\nu}}{\rm MeV}\right) \left( \frac{r}{100 {\rm
  km}}\right)^2 \left( \frac{\langle \sigma _{\nu} \rangle}{10^{-41}
  {\rm cm}^2} \right)^{-1} s
\label{eqn:tnu}
\end{eqnarray}
where $L_{\nu, 51}$ and $\epsilon _{\nu}$ have already been defined in
Sec.~2-1, and $\langle \sigma _{\nu} \rangle$ is the averaged cross
section over neutrino energy spectrum. 
As discussed above, neutrino heating occurs most effectively at $r
\approx 12$ km (see also Fig.~\ref{fig5}(a)), and we set this value in
Eq.~(\ref{eqn:tnu}). Since two neutrino processes (\ref{eqn:mib}) and
(\ref{eqn:plb}) make the biggest contribution to heating the wind and
$\epsilon _{\nu_e}=12$ MeV and $\epsilon _{\bar{\nu}_e}=22$ MeV, we set 
$\epsilon _{\nu}=(\epsilon _{\nu_ e} + \epsilon _{\bar{\nu}_e} )/2
\approx 15$ MeV.
We take $\langle \sigma _{\nu} \rangle = 10^{-41}$ cm$^2$. Incorporating 
these values into Eq.~(\ref{eqn:tnu}), we can obtain $\tau_{\nu}$
value.
Let us compare the specific collision time, $\tau_{\nu}$, and the
proper expansion time, $\tau_{\rm heat}$, with each other:
\begin{mathletters}
\begin{eqnarray}
\tau_{\nu}&=&0.0043 {\rm s} ~~< ~~ \tau_{\rm heat}= 0.017~~{\rm s},
\nonumber
\\
&&{\rm  for}~~(L_{\nu},M)=(10^{52} {\rm ergs/s}, 2.0
  M_{\odot}), 
\nonumber
\\
&&
\label{eqn:taua} \\
\tau_{\nu}&=& 0.043 {\rm s} ~~< ~~  \tau_{\rm heat}= 0.28 {\rm s},
\nonumber
\\
&&{\rm for}~~(L_{\nu},M)=(10^{51} {\rm ergs/s}, 1.4
  M_{\odot}).
\nonumber
\\
&&   
\end{eqnarray}
\end{mathletters} 
We can conclude that there is enough time for the expanding wind to be
heated by neutrinos even with short dynamic time scale
for the $\alpha$-process, $\tau_{\rm dyn} \sim 6$ ms, which
corresponds to the case (\ref{eqn:taua}).

Before closing this section, let us briefly discuss the effect of
electron fraction $Y_e$ on the hydrodynamic condition of the
neutrino-driven wind.
Although we took $Y_e=0.5$ for simplicity
in our numerical calculations, we should examine the sensitivity of the
calculated result on $Y_e$ quantitatively.
Since we are interested in short dynamic time scale, let us investigate
the case with $(L_{\nu},M)=(10^{52} {\rm ergs/s}, 2.0 M_{\odot})$ which
results in $S=138.5$ and $\tau_{\rm dyn}=0.00618$ s for $Y_e = 0.5$.
When we adopt $Y_e = 0.4$, these quantities change slightly to $S=141.5$ 
and $\tau_{\rm dyn}=0.00652$ sec.
These are very small changes less than $5 \%$, and the situation in
similar for the other sets of $(L_{\nu},M)$.

To summarize this section, we found that there is a parameter 
region in Fig.~\ref{fig8} which leads to desirable physical condition for the
r-process nucleosynthesis.
Sophisticated supernova simulation (\cite{wil}) indicates that the
neutrino luminosity from the protoneutron star decreases slowly from
about $5 \times 10^{52}$ to $10^{51}$ ergs/s as the time passes by after 
the core bounce. 
Therefore, our favorable neutrino luminosity $L_{\nu}=10^{52}$ ergs/s is 
possible in reality in relatively earlier epoch of supernova explosion at 
around 0.5 s to a few seconds after the core bounce.
        
\section{R-process nucleosynthesis calculation}

Our discussion on the r-process nucleosynthesis in the last section was
based on  Hoffman's criterion, Eq.~(\ref{eqn:hofcon}), which is to be
referred with caution for several assumptions and approximations
adopted in its derivation.
The purpose of this section is to confirm quantitatively that the
r-process occurs in the neutrino-driven wind with short 
dynamic time scale, which we found in the present study.

Given the flow trajectory characterized by $u(t)$, $\rho_b(t)$, and $T(t)$ 
as discussed in the last section, our nucleosynthesis calculation starts 
from the time when the temperature is $T_9=9$.
Since this temperature is high enough for the system to be in the NSE,
initial nuclear composition consists of free neutrons and protons.
We set $Y_e=0.4$ in order to compare with Hoffman's criterion shown in
Fig.~\ref{fig8}.
In our nucleosynthesis calculation we used a fully implicit single
network code for the $\alpha$-process and r-process
including over 3000 isotopes.
We take the thermonuclear reaction rates for all relevant nuclear processes
and their inverse reactions
as well as 
weak interactions from Thielemann~(1995)
for the isotopes $Z \leq 46$ and from Cowan et al.~(1991) for the
isotopes $Z > 46$.
Previous r-process calculations had complexity that the 
seed abundance distribution at $T_9=2.5$ was not fully shown in literature
(\cite{wil,wh,hof}), 
which makes the interpretation of the whole nucleosynthesis process less 
transparent.
This inconvenience happened because it was numerically too heavy to run
both $\alpha$-process and r-process in a single network code for huge
number of reaction couplings among $\sim 3000$ isotopes.
For this reason, one had to calculate the $\alpha$-process first, using
smaller network for light-to-intermediate mass elements, in order to
provide seed abundance distribution at $T_9 = 2.5~(T \approx 0.2 {\rm MeV})$.
Adopting such seed abundance distribution and following the evolution of 
material in the wind after $T \approx 0.2$ MeV, which is the onset
temperature of the r-process, the r-process nucleosynthesis calculation
was extensively carried out by using another network code independent
of the $\alpha$-process.
Our nucleosynthesis calculation is completely free from this complexity
because we exploited single network code which 
is applied to a sequence of the whole processes of NSE - $\alpha$-process -
r-process.

The calculated mass abundance distribution is shown in Figs.~\ref{fig10}
and \ref{fig9} for 
the neutrino-driven wind with $(L_{\nu},M)=(10^{52}ergs/s, 2.0
M_{\odot})$ that makes most favorable condition for the r-process
nucleosynthesis with the shortest $\tau_{\rm dyn}=0.0062$ s among those
studied in the present paper~(see Fig.~\ref{fig8}).
Figure~\ref{fig9} displays the snapshot at the time when the
temperature cooled 
to $T_9 = 2.5~(\approx 0.2 {\rm MeV})$ at the end of the
$\alpha$-process.
This shows seed abundance distribution at the onset of the
r-process, too.
Our calculated quantities at this temperature are the baryon mass
density $\rho _{\rm b} =3.73 \times 10^4 $ g/cm$^3$, neutron mass
fraction $X_n = 0.159$, mass fraction of alpha-particle
$X_{\alpha}=0.693$, average mass number of seed nuclei $\langle A \rangle
=94$, and neutron-to-seed abundance ratio $n/s = 99.8$ for the set of
hydrodynamic quantities $\tau_{\rm dyn}= 0.0062$ s, $S \approx 139$, and 
$Y_e=0.4$.
These values should be compared with those adopted in Woosley's
calculation of trajectory 40, {\it i,e,} $\rho_b =
1.107 \times 10^4$ g/cm$^3$, $X_n = 0.176$, $X_{\alpha}=0.606$, $\langle 
A \rangle = 95$, $n/s=77$, $\tau_{\rm dyn} \approx 0.305$ s, $S
\approx 400$, and $Y_e=0.3835$, as in Table 3 in Woosley et al.~(1994).
It is interesting to point out that our seed abundance distribution in
Fig.~\ref{fig9} is very similar to theirs (\cite{wil,wh}), as clearly shown by
almost the same  $\langle A \rangle \approx  95$, although the other
evolutionary parameters and thermodynamic quantities are different from
each other. The calculated final r-process abundance is displayed in
Fig.~\ref{fig10}.
Our wind model can produce the second $(A \approx 135)$ and third $(A
\approx 195)$ r-process abundance peaks and rare earth elements between
them as well.  

It is generally accepted that the r-process elements will be produced if 
there are plenty of free neutrons and if the neutron-to-seed abundance
ratio is high enough to approximately satisfy $A \approx \langle A
\rangle +n/s$ 
(\cite{hof}) at the beginning of the r-process,
where $A$ is the typical mass number of the r-process element.  
Therefore, the $\alpha$-process should take the key to understand 
why our wind model results in a similar r-process nucleosynthesis to the
result of Woosley's 
trajectory 40.

The $\alpha$ burning starts when the temperature cools below $T = 0.5$
MeV.
Since triple alpha reaction $^4$He$(\alpha \alpha,\gamma)^{12}$C is too
slow at this temperature, alternative nuclear reaction path to reach $^{12}$C,
$^4$He$(\alpha n,\gamma)$ 
\\
$^9$Be$(\alpha,n)^{12}$C, triggers explosive
$\alpha$-process to 
produce the seed elements.
In rapidly expanding flow of neutrino-driven wind with short $\tau_{\rm
  dyn}$, it is not a good approximation to assume that the first reaction
$^4$He$(\alpha n, \gamma)^9$Be is in the NSE.
Rate equation is thus written as 
\begin{eqnarray}
\frac{d Y_9}{dt} &\approx& \rho_b^2 Y_{\alpha}^2 Y_n \lambda(\alpha \alpha n 
\rightarrow ^9{\rm Be})
\nonumber
\\ 
&-& \rho_b Y_{\alpha}Y_9 \lambda(^9{\rm Be}~ \alpha
\rightarrow ^{12}{\rm C})
\nonumber\\
&+ &(their~~ inverse~~and~~
\nonumber
\\
&&other~~reaction~~ rates),
\label{eqn:rate}
\end{eqnarray}
where $Y_9$, $Y_{\alpha}$, $Y_n$ are the number fractions of $^9$Be,
$\alpha$, and neutron, and $  \lambda(\alpha \alpha n 
\rightarrow ^9{\rm Be})$ and $ \lambda(^9{\rm Be} \alpha
\rightarrow ^{12}{\rm C})$ are the thermonuclear reaction rate for each
reaction process as indicated.
Details on $\lambda$'s are reported in Woosley and Hoffman (1992) and
Wrean, Brune, and Kavanagh (1994).
Let us take the first term of the {\it r.h.s.} of Eq.~(\ref{eqn:rate})
which is largest all terms in Eq.~(\ref{eqn:rate}).
This is allowed in the following discussion of the time scale because
the $^4$He$(\alpha n,\gamma)^9$Be reaction is the slowest among all
charged particle reaction paths in all $\alpha$-process reactions.
We now define the typical nuclear reaction time scale $\tau_{\alpha}$ of 
the $\alpha$-process, regulated by the  $^4$He$(\alpha n,\gamma)^9$Be
reaction time scale $\tau_N$, as
\begin{equation}
\tau_{\alpha} \gtrsim \left(\rho_b^2 Y_{\alpha}^2 Y_n \lambda(\alpha \alpha n 
\rightarrow ^9{\rm Be}) \right)^{-1} \equiv \tau_{N}.
\label{eqn:35}
\end{equation}
We show the ratio  $\tau_{dyn}/\tau_N$ as a function of the baryon mass
density $\rho_b$ at the beginning of the $\alpha$-process when $T=0.5$
MeV for various cases of the wind models with $(L_{\nu}, M)$ in
Fig.~\ref{fig11}. 
Note that the critical line  $\tau_{dyn}/\tau_{\alpha} =1$ is slightly
shifted upwards 
because of $\tau_{N} \lesssim \tau_{\alpha}$.
This figure, with the help of Fig.~\ref{fig8}, clearly indicates that the
favorable conditions for the r-process nucleosynthesis have inevitably
shorter $\tau_{\rm dyn} \ll \tau_{N}$ and $\tau_{\alpha}$.
Typical ratio is of order $\tau_{dyn}/\tau_N \sim 0.1$.
To interpret this result, there is not enough time for the
$\alpha$-process to accumulate a number of seed elements and plenty of
free neutrons are left even at the beginning of 
the r-process. Consequently, the $n/s$ ratio becomes very high $\sim 100$.

As for the neutron mass fraction, on the other hand, our value $X_n =
0.159$ is smaller than Woosley's model value $X_n=0.176$ in trajectory
40 because low entropy is in favor of low neutron fraction.
This may be a defect in our low entropy model.
However, the short dynamic time scale saves the situation by regulating
the excess of the seed elements as discussed above.
These two effects compensate with each other to result in average mass
number of seed nuclei $\langle A \rangle \approx 95$ and neutron-to-seed
abundance ratio $n/s \approx 100$, which is ideal for the production of
the third $(A \approx 195)$ abundance peak of the r-process 
elements in our model, as displayed in Fig.~\ref{fig10}. 

R-process elements have recently been detected in several
metal-deficient halo stars ~(\cite{sn}) and the relative abundance
pattern for the elements between the second and the third peak proves to 
be very similar to that of the solar system r-process abundances.
One of the possible and straightforward interpretations of this fact is
that they were produced in narrow window of some limited
physical condition in massive supernova explosions, as studied in the
present paper. 
These massive stars have short lives $\sim 10^7$ yr and eject
nucleosynthesis products into interstellar medium continuously from the
early epoch of Galaxy evolution.
It is not meaningless, therefore, to discuss several features of our
calculated result in comparison with the solar system r-process
abundance distribution (\cite{kap}) in Fig.~\ref{fig10}.  
Although K\"appeler et al. obtained these abundances as s-process
subtractions from the observed meteoritic abundances~(\cite{and}) for the 
mass region $63 \leq A \leq 209$, the inferred yields and error bars for 
$A=206,207,208$, and 209 are subject to still uncertain s-process
contribution.
We did not show these heavy elements $A=206-209$ in Fig.~\ref{fig10}.

Our single wind model reproduces observed abundance peaks around $A
\approx 130$ and $A \approx 195$ and the rare earth element region
between these 
two peaks.
However, there are several requirements to the wind model in order to
better fit the details of the solar system r-process abundances in the mass
region $120 \lesssim A$.
The first unsatisfactory feature in our model calculation is that the
two peaks are shifted upward by $2 \sim 4$ mass unit, although overall
positions and peaks are in good agreement with the solar system data.
This is a common problem in all theoretical calculations of the
r-process nucleosynthesis (\cite{mey3,wil}).
The shift of the peak around $A\approx195$ is slightly larger than that
around at $A \approx 130$, which may be attributed to a strong neutron
exposure as represented by $n/s \approx 100$ in our model calculation.
The second feature is that the rare earth element region shows broad
abundance hill, but its peak position $A \approx 165$ in the data is not 
explained in our calculation.
It was pointed out by Surman et al.~(1997) that the abundance structure
in this mass region is sensitive to a subtle interplay of nuclear
deformation and beta decay just prior to the
freeze-out of the r-process.
More careful studies of these nuclear effects and the dynamics of the
r-process nucleosynthesis are desirable.
The third failure in the model calculation is the depletion around
$A\approx120$, which is also another serious problem encountered
commonly by all previous theoretical calculations.
This deficiency is thought to be made by too fast runaway of the
neutron-capture reaction flow in this mass region.
This is due to too strong shell effects of the $N=82$ neutron shell
closure, suggesting an incomplete nuclear mass extrapolations to the
nuclei with $Z \approx 40$ and $N \approx 70-80$ which correspond to the 
depleted abundance mass region $A \approx 120$.
It is an interesting suggestion ~among many others
(\cite{wil}) that an artificial smoothing of extrapolated zigzag
structure of nuclear masses could fill the abundance dip around $A
\approx 120$.
This suggestion sheds light on the improvement of mass formula.

Let us repeat it again that an overall success in the present r-process
nucleosynthesis calculation, except for several unsatisfactory fine
features mentioned above, is only for heavier mass elements $130
\lesssim A$ including the second $(A \approx 130)$ and the third $(A
\approx 195)$ peaks.
When one looks at disagreement of the abundance yields around the first
$(A \approx 80)$ peak, relative to those at the third peak, between our
calculated result and the solar system r-process abundances, it is clear 
that a single wind model is unable to reproduce all three r-process
abundance peaks.
The first peak elements should be produced on different conditions with
lower neutron-to-seed ratio and higher neutrino flux.
It has already been pointed out by several authors (\cite{see,kod,hil}) 
that even the r-process nucleosynthesis needs different neutron
exposures similarly to the s-process nucleosynthesis in order to
understand the solar system r-process abundance distribution.
In a single event of supernova explosion, there are several different
hydrodynamic conditions in different mass shells of the neutrino-driven
wind (\cite{wil,jan1}), which may produce the first peak elements.
Different progenitor mass supernova or the event like an exploding
accretion disk of neutron-star merger might contribute to the
production of the r-process elements.
Consideration of these possibilities is beyond our scope in the present paper. 

We did not include the effects of neutrino absorption and scattering
during the nucleosynthesis process in the present calculation. This is because
these effects do not change drastically 
the final r-process yields as far as the dynamic expansion time scale
$\tau _{\rm dyn}$ is very short.
Using Eq.~(\ref{eqn:tnu}), we can estimate the specific collision time
for neutrino-nucleus interaction  
\begin{equation}
\tau_{\nu} \approx 0.082 {\rm s} - 0.31 {\rm s},
\label{eqn:34}
\end{equation}
where the input parameters are set equal to $L_{\nu ,51}=10,
\epsilon_{\nu}=15$ MeV, and $\langle \sigma _{\nu} \rangle = 10^{-41}$
cm$^2$.
Note that $\tau_{\nu} \approx 0.082$ s is the specific neutrino
collision time at $r=52$ km where the temperature of the wind becomes
$T=0.5$ MeV at the beginning of the $\alpha$-process, and $\tau_{\nu}
\approx 0.31$ s for $r=101$ km and $T=0.5/e \approx 0.2$ MeV at the
beginning of the r-process.
These $\tau_{\nu}$ values are larger than $\tau_{\rm dyn}=0.0062$ s that
stands for the duration of the $\alpha$-process by definition.
Therefore, the neutrino process does not disturb the hydrodynamic condition of 
the rapid expansion during the $\alpha$-process.

It is to be noted, however, that the neutrino process virtually makes
the strong effect on the r-process for the winds of slow expansion.
We have numerically examined Woosley's model~(1994) of trajectory 40 to
find $\tau_{\rm dyn} \approx 0.3$ s.
Meyer et al. ~(1998) also used $\tau_{\rm dyn}=0.3$ s in their
simplified fluid trajectory to see the neutrino-capture effects.
This dynamic time scale $\tau_{\rm dyn} \approx 0.3$ s is
larger than or comparable to the specific neutrino collision time
$\tau_{\nu}$ in Eq.(\ref{eqn:34}).  
In such a slow expansion the neutrino absorption
by neutron~(\ref{eqn:mib}) proceeds to make a new proton in the
$\alpha$-process. 
This proton is quickly interconverted into alpha-particle in the
following reaction chain, $p(n,\gamma)d(n,\gamma)$t which is followed by
t$(p,n)^3$He$(n,\gamma)^4$He and t$(t,2n)^4$He, and contributes to the
production of seed elements.
These radiative capture reactions and nuclear reactions are much faster
than the weak process (\ref{eqn:plb}) on newly produced proton from the
process (\ref{eqn:mib}).
The net effect of these neutrino processes, therefore, is to decrease the
neutron number density and increase the seed abundance, which leads to
extremely low $n/s$ ratio.
As a result, even the second abundance $(A \approx 130)$ peak of the
r-process elements 
disappears, as reported in literature (\cite{mey,mey2}).
Details on the neutrino process will be reported elsewhere.

We have assumed that electrons and positrons are fully relativistic
throughout the nucleosynthesis process.
However, the total entropy of the system may change at the temperature
$T \lesssim 1/3 m_e$ where electrons and positrons tend to behave as
non-relativistic particles.
This might affect the nucleosynthesis although it does
not affect significantly the dynamics near the protoneutron star.
We should correct this assumption in the future papers.

Finally, let us refer to massive neutron star.
Large dispersion in heavy element abundance of halo stars has recently
been observed.
Ishimaru and Wanajo (1999) have shown in their galactic chemical
evolution model that if r-process nucleosynthesis occurs in either
massive supernovae $\geq 30 M_{\odot}$ or small mass supernovae $8-10
M_{\odot}$, where these masses are for the progenitors, the observed
large dispersion can be well explained theoretically.
In addition, SN1994W and SN1997D are presumed to be due to 25
$M_{\odot}-40 M_{\odot}$ massive progenitors because of very low $^{56}$Ni 
abundance in the ejecta (\cite{SN1,SN2}).
These massive supernova are known to have massive iron core $\geq 1.8
M_{\odot}$ and leave massive remnant (\cite{SN2}).
It is critical for the r-process nucleosynthesis whether the remnant is
neutron star or black hole.
Recent theoretical studies of the EOS of neutron star matter, which is
based on relativistic mean field theory, set upper limit of the neutron star 
mass at 2.2 $M_{\odot}$ (\cite{shen}).

\section{Conclusion and Discussions}

We studied the general relativistic effects on neutrino-driven
wind which is presumed to be the most promising site for the
r-process nucleosynthesis.
We assumed the spherically symmetric and steady state flow of the wind.
 In solving the basic equations for relativistic fluid in Schwarzschild
geometry, we did not take approximate method as adopted in several
previous studies.  
We tried to extract generic properties of the wind in
manners independent of supernova models
or neutron-star cooling models. 

The general relativistic effects introduce several 
corrections to the equations of motion of the fluid and also to 
the formulae of neutrino heating rate due to the
redshift and bending of neutrino trajectory.
We found that these corrections increase
entropy and decrease dynamic time scale of the expanding neutrino-driven
wind from those in the Newtonian case.
The most important corrections among them proves to be the correction to
the hydrodynamic equations. Both temperature and density of the
relativistic wind decrease more rapidly than the Newtonian wind as the
distance increases without
remarkable change of the velocity at $r<30$km, where the neutrino-heating 
takes place efficiently.
The lower the temperature and density are, the larger the net heating rate is.
This is the main
reason why the entropy in the relativistic case is larger than the
Newtonian case. 

We also looked for suitable environmental condition for the r-process
nucleosynthesis in general relativistic framework.
We studied first the differences and similarities between
relativistic and Newtonian winds in numerical calculations, and then
tried to interpret their behavior by expressing gradients of the
temperature, velocity and density of the system analytically under the
reasonable approximations. 
We extensively studied the key quantities for the
nucleosynthesis, {\it i.e.} the entropy $S$ and the dynamic time scale
$\tau_{dyn}$ of the expanding neutrino-driven wind, and their dependence 
on the protoneutron
star mass, radius, and neutrino luminosity.
We found that more massive or equivalently more compact neutron star
tends to produce explosive neutrino-driven wind of shorter dynamic
time scale, which is completely
different from the result of the previous studies in the Newtonian case
which adopted approximation methods. 
We also found that the entropy becomes larger as the neutron star mass
becomes larger.
Since the larger luminosity makes the dynamic time scale shorter, the large
neutrino luminosity is desirable as far as it is less than
$10^{52}$ergs/sec.  
If it exceeds $10^{52}$ ergs/sec, only the mass outflow rate becomes large 
and the flow can not cool down to $\sim 0.2 \rm{MeV}$
by the shock front $r\sim 10,000$km. 
As the result, the time scale becomes too long, which is not favorable to 
the r-process nucleosynthesis.

Although we could not find a model which produces very large
entropy $S\sim$ 400 as suggested by Woosley et al.(1994), it does not
mean that the r-process does not occur in the neutrino-driven wind.
We compared our results with Hoffman's condition and found that the short
dynamic time scale $\tau_{dyn}\sim 6$ms, with $M=2.0 M_{\odot}$ and
$L_{\nu}=10^{52}$ ergs/sec, is one of the most preferable condition for
producing r-process elements around the third peak ($A \sim 195$).
In order to confirm this,
we carried out numerical calculations of the r-process nucleosynthesis
upon this condition by using fully implicit single network code which takes
account of more than $\sim 3000$ isotopes and their associated nuclear
reactions in large network.
We found that the r-process elements around $A \sim 195$ and even the 
heavier elements like thorium can be produced in this wind,
although it has low entropy $S \sim 130$.
The short dynamic time scale $\tau_{dyn}\sim 6$ ms was found to play the
role so that the few seed nuclei are produced with plenty of free neutrons
left over at the beginning of the r-process.
For this reason the resultant neutron-to-seed ratio, $n/s \sim 100$,
in high enough even with low entropy and leads to appreciable production 
of r-process elements around the second$(A \approx 130)$ and third $(A
\approx 195)$ abundance peaks and even the hill of rare earth elements between
the 
peaks.

Note that the energy release by the interconversion of nucleons into
$\alpha$-particles at $T\sim 0.5$ MeV produces an additional entropy
about $\Delta S\sim 14$. 
This was not included in our present calculation. 
We can make note that, taking account of this increase, the r-process
could occur in the neutrino-driven wind from hot neutron star 
whose mass is smaller than 2.0$M_\odot$.

One might think that short $\tau_{dyn}$ brings
deficiency of neutrino heating and that the wind may not blow.
It is not true because the mass elements in the wind are heated by
energetic neutrinos most efficiently at $r \lesssim30km$, while the
expansion time scale $\tau_{dyn}$ is
the time for the temperature to decrease from $T\sim 0.5$ MeV to 0.2 MeV
at larger radii.
The duration of time for the mass elements to reach 30km after leaving
neutron star surface is longer than $\tau_{dyn}$. There is enough time
for the system to be heated by neutrinos even for $\tau_{dyn}$ as low as
$\sim 6$ ms. 

We did not include neutrino-capture reactions that may change $Y_e$
during the nucleosynthesis process.
Since the initial electron fraction was taken to be relatively high $Y_e 
=0.4 \sim 0.5$, there is a possibility that the final nucleosynthesis
yields in neutrino-driven wind may be modified by the change in $Y_e$
during the $\alpha$- and r-processes.
However, this is expected to make a small modification in our present expansion
model with short dynamic time scale because the typical time scale of
neutrino interaction is longer than $\tau_{dyn}$.  
We will report the details about the nucleosynthesis
calculation including neutrino-capture reactions in forthcoming papers.  

It was found that the entropy decreases with increasing neutrino
luminosity. This fact  
suggests that one cannot obtain large entropy
by merely making the heating rate large.
The cooling rate, on the other hand, does not depend on the neutrino
luminosity. In the present studies we included two cooling mechanisms of 
the $e^+ e^-$ capture by free nucleons and the $e^+ e^-$ pair
annihilation. As for the cooling rate due to the $e^+ e^-$ pair
annihilation, only the contribution from pair-neutrino process is
usually taken into consideration, as in the present calculation.
However, there are many other processes which can contribute to the total
cooling rate.
They are the photo-neutrino process, the plasma-neutrino process, the
bremsstrahlung-neutrino process and the recombination-neutrino process
(\cite{ito}).  
Indeed, if we double our adopted cooling rate artificially, we can
obtain larger entropy.
Details on the numerical studies of the cooling rate are reported
elsewhere.
The radial dependence of the heating
rate is also important (\cite{qw}).
Since both heating and cooling processes are critical to
determine the entropy, more investigation on the neutrino
process is desirable.

There are other effects which have not been included in the present study.
They are, for example, the mass accretion onto the neutron star, the
time variation of the neutrino luminosity, convection and mixing of
materials, and rotation or other dynamic process which break spherical
symmetry of the system.
These probably important effects may make several modifications to the
present result.
However, we believe that our main conclusion that there is a possibility 
of finding the r-process nucleosynthesis in an environment of
relatively small entropy and short dynamic time scale is still valid.
We conclude that the neutrino-driven wind is a promising astrophysical site for
the successful r-process nucleosynthesis. 

\acknowledgments
We are grateful to Prof.~G.J.~Mathews and Prof.~J.~Wilson for many
useful discussions and kind advice.
We also would like to thank Profs.~R.N.~Boyd, S.E.~Woosley, H.~Toki,
Drs.~K.~Sumiyoshi, S.~Yamada, and H.~Suzuki for their stimulating discussions.
This work has been supported in part by the Grant-in-Aid for Scientific
Research (1064236, 10044103) of the Ministry of Education, Science,
Sports and Culture of Japan and the Japan Society for the Promotion of Science.

\clearpage

\figcaption[f1.eps]{\footnotesize
Outflow velocity (a) and temperature (b) in Schwarzschild geometry as a
function of the distance $r$ 
from the center of the neutron star for various mass outflow rate
$\dot{M}$, where neutron star mass $M=1.4 M_{\odot}$ and neutrino
luminosity $L_{\nu_e}=10^{51}$ ergs/s are used.
Long dashed curve is for the critical mass outflow rate $\dot{M}_{\rm
crit} = 5.2681 \times 10^{-6}M_{\odot}$, in which the velocity becomes
supersonic through the critical point.
Fives curves denoted by 1 to 5 corresponds respectively to  
$\dot{M}=5.25 \times 10^{-6}$, $5.15 \times
10^{-6}$, $5.0855 \times 10^{-6}$, $5.0\times 10^{-6}$, and $4.8 \times
10^{-6}~M_{\odot}$.
Calculated result denoted by ``3'' meets with our imposed boundary
condition of $T=0.1$ MeV at $r=10000$ km.
Entropy per baryon $S$ and dynamic timescale $\tau_{\rm dyn}$, which
correspond to each curves from 1 to 5, are tabulated in Table 1. 
Note that the temperature denoted by ``5'' does not decrease
to T = 0.5/e MeV within 10000 km (see Table 1).
\label{fig1}}

\figcaption[f2.eps]{\footnotesize
The same as those in Fig.1, for the case of $M=2.0M_{\odot}$,
$L_{\nu_e}=10^{52}$ ergs/s.
Long dashed curve is for the critical mass outflow rate $\dot{M}_{\rm
crit} = 1.2459 \times 10^{-4}M_{\odot}$, 
and fives curves denoted by 1 to 5 correspond respectively to  
$\dot{M}=1.245 \times 10^{-4}$, $1.240 \times
10^{-4}$, $1.225 \times 10^{-4}$, $1.215\times 10^{-4}$, and $1.195 \times
10^{-4}~M_{\odot}$.
Calculated result denoted by ``1'' meets with our imposed boundary
condition of $T=0.1$ MeV at $r=10000$ km.
Entropy per baryon $S$ and dynamic timescale $\tau_{\rm dyn}$, which
correspond to each curves from 1 to 5, are tabulated in Table 1. 
Note that the temperature denoted by ``5'' does not decrease
to T = 0.5/e MeV within 10000 km (see Table 1).
\label{fig2}} 

\figcaption[f3.eps]{\footnotesize
Outflow velocity $u(r)$, in unit of $10^7\rm{cm/sec}$, temperature
$T(r)$, in unit of 0.1MeV, and baryon mass density $\rho_b(r)$, in unit
of $10^8g/cm^3$, as function of the distance $r$ from the center of the
neutron star with the protoneutron star mass $M = 1.4{\rm M_{\odot}}$,
radius $R = 10$ km, neutrino luminosity $L_{\nu}=10^{51}$ ergs/s and
initial density $10^{10}{\rm g/cm}^3$. 
Solid and broken lines display the results in Schwarzschild and
Newtonian geometries, respectively. 
We choose the mass outflow rate $\dot{\rm{M}}=5.0855 \times
10^{-6}\rm{M}_{\odot}$/s for the Schwarzschild case and
$\dot{\rm{M}}= 1.2690 \times 10^{-5}\rm{M}_{\odot}$ for the  Newtonian
case. See text for details of the outer boundary condition on $\dot{M}$.
\label{fig3}}

\figcaption[f4.eps]{\footnotesize
Entropy per baryon $S(r)$ as a function of the distance $r$ from the
canter of the neutron star. Solid and broken lines are the same as those 
in Fig.~1 for the same set of the input parameters.
\label{fig4}}

\figcaption[f5.eps]{\footnotesize
Specific neutrino heating rate $\dot{q}(r)$ as a function of the
distance $r$ from the center of the neutron star, for the same set of
the input parameters as those in Fig.~1.   
(a)~Total net heating rate $\dot{q}$. The solid and broken lines are for 
the Schwarzschild (denoted by general relativity) and Newtonian case,
respectively. 
(b)~Decomposition of the net heating rate into five different
contributions from the heating process $\dot{q}_1$, $\dot{q}_3$, and
$\dot{q}_5$ (solid lines) and the cooling process $\dot{q}_2$ and
$\dot{q}_4$ (broken line) for the Schwarzschild case. See text for
details of $\dot{q}_i$.
(c)~The same as those in (b) for the Newtonian case.
\label{fig5}}

\figcaption[f6.eps]{\footnotesize
Dynamic time scale $\tau_{dyn}$ (a) and entropy per baryon $\rm{S}$ (b)
vs. neutron star mass $M$ at 0.5MeV.
Closed circles, connected by thick solid line, and open triangles,
connected thin solid line, are the calculated results for the
Schwarzschild and Newtonian case, respectively, by using the same
set of the input parameters as in Fig.~1.
Two broken lines are from Qian and Woosley (1996) in Newtonian case,
which adopted an assumption of the radiation dominance (lower in
$\tau_{\rm dyn}$, and upper in $S$) or the dominance of non-relativistic 
nucleon (upper in $\tau_{\rm dyn}$, and lower in $S$).
\label{fig6}} 

\figcaption[f7.eps]{\footnotesize
Dynamic time scale $\tau_{dyn}$ (a) and entropy per baryon $\rm{S}$ (b)
vs. neutrino luminosity $L_{\nu}$ at $T=0.5$ MeV.
Thick and thin lines and two broken lines are the same as those in
Fig.~4.
At the larger and of $L_{\nu} \sim 10^{52}$ ergs/s, there is no solution 
to satisfy our imposed boundary condition, $T=0.1$ MeV at $r=10000$ km.
See text details. \label{fig7}}  

\figcaption[f8.eps]{\footnotesize
Relation between entropy per baryon S and dynamic time scale
$\tau_{dyn}$ for various combinations of the neutron star mass
$1.2M_{\odot} \leq M \leq 2.0M_{\odot}$ and the 
neutrino luminosity $10^{50} \leq L_{\nu} \leq 10^{52}$ ergs/s.
Solid and broken lines connect the same mass and luminosity.
At the largest end of $L_{\nu} \sim 10^{52}$ ergs/s for each $M$,
there is no solution to satisfy our imposed boundary condition, $T=0.1$ MeV at
$r=10000$ km.
Two zones indicated by shadows  satisfy the approximate conditions, for
$Y_e=0.4$, on which the successful r-process occurs (Hoffman et
al. 1997) to make  
the second abundance peak around $A=130$ (lower) and the third abundance 
peak around $A=195$ (upper).
See text for details.
\label{fig8}}

\figcaption[f9.eps]{\footnotesize
Seed abundances at $T_9=2.5$ as a function of atomic number $A$.
See text for details. \label{fig9}}

\figcaption[f10.eps]{\footnotesize
Final r-process abundances (lines) as a function of atomic mass number
$A$ compared with the solar system r-process abundances (filled circles) 
from K\"appeler, Beer, \& Wisshak (1989).
The solar system r-process abundances are shown in arbitrary unit. See
text for details. 
\label{fig10}}

\figcaption[f11.eps]{\footnotesize
The ratio of dynamic time scale $\tau_{\rm dyn}$ to 
the time scale of typical $\alpha$-process nuclear reaction $\tau_{\rm N}$,
$\tau_{\rm dyn}/\tau_{\rm N}$, vs. baryon  mass density at $T=0.5$ MeV,
for various combinations of the neutron star mass $1.2 M_{\odot} \leq M
\leq 2.0 M_{\odot}$ and the neutrino luminosity $10^{50} \leq L_{\nu}
\leq 10^{52}$  ergs/s.
Solid and broken lines connect the same mass and luminosity.
\label{fig11}}

\clearpage

\begin{deluxetable}{ccccc}
\tablecaption{Entropy and dynamic time scale for different $\dot{M}$.
Since the temperature in the 5th case for both  $M=1.4M_{\odot}$ and
$M=2.0M_{\odot}$ 
does not decrease to T = 0.5/e MeV within 10000 km,
$\tau_{\rm dyn}$ is not defined (see Fig. 1(b) and 2(b)).
}
\tablehead{
&\colhead{}&\colhead{$\dot{M}$} & \colhead{entropy} &
\colhead{$\tau_{\rm dyn}$} \\
&\colhead{}&\colhead{$(10^{-6}M_{\odot}/sec)$} & \colhead{$(k)$} &
\colhead{$(s)$}} 

\startdata
1.4$M_{\odot},10^{51}$ergs/s&
$M_{crit}$&5.2681 & 116 & 0.037171\\
&1&5.2500 & 117 & 0.041304\\
&2&5.1500 & 120 & 0.084335\\
&3&5.0855 & 123 & 0.16455\\
&4&5.0000 & 126& 0.71569\\
&5&4.8000 & 135 & \\
\tableline
2.0$M_{\odot},10^{52}$ergs/s&
$M_{crit}$&$1.2459 \times 10^{2}$ & 138& 0.00507\\
&1&$1.2450 \times 10^{2}$ & 138 & 0.00618\\
&2& $1.2400\times 10^{2}$& 139 &$ 0.01088$\\
&3&$1.2250\times 10^{2}$ & 141 & $0.08962$\\
&4&$1.2150\times 10^{2}$ & 143 & 2.6272\\
&5&$1.1950\times 10^{2}$ &146  & 
\enddata
\end{deluxetable}

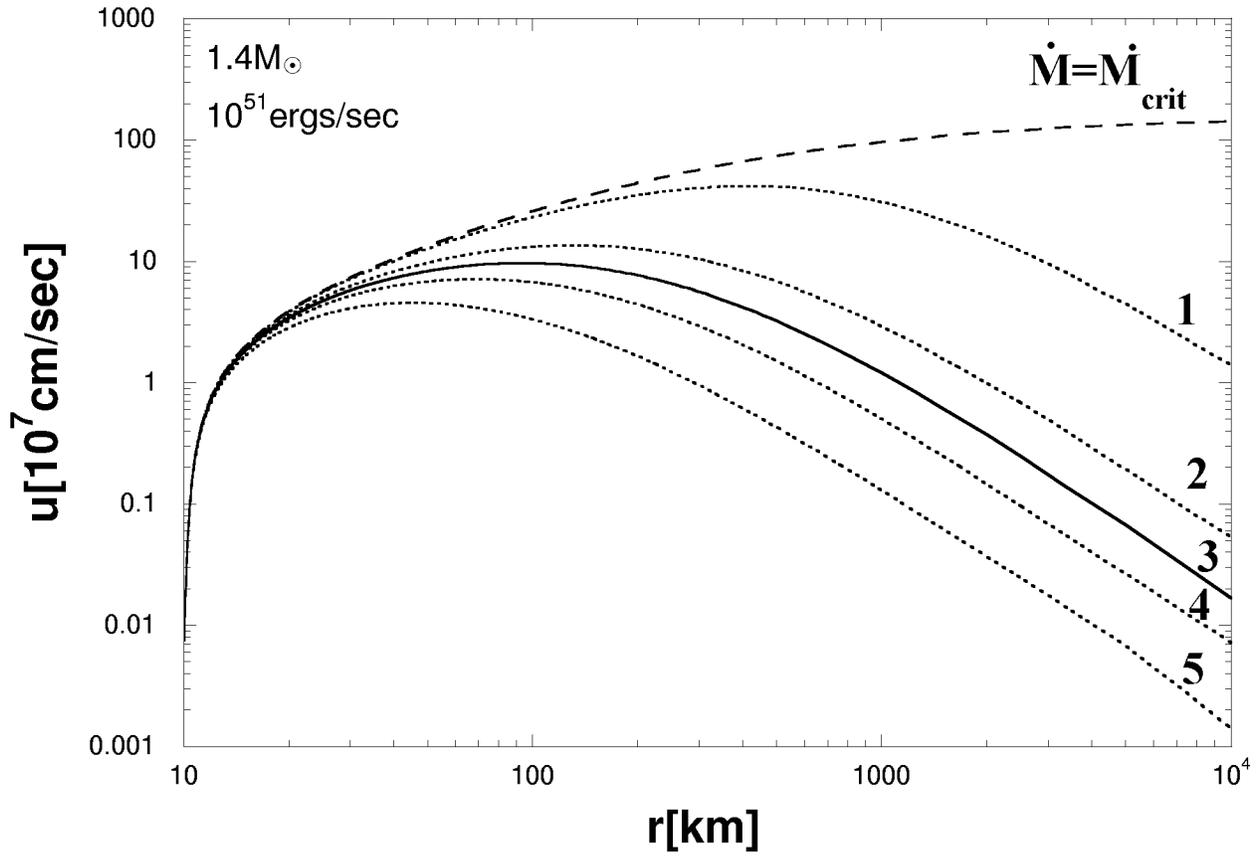
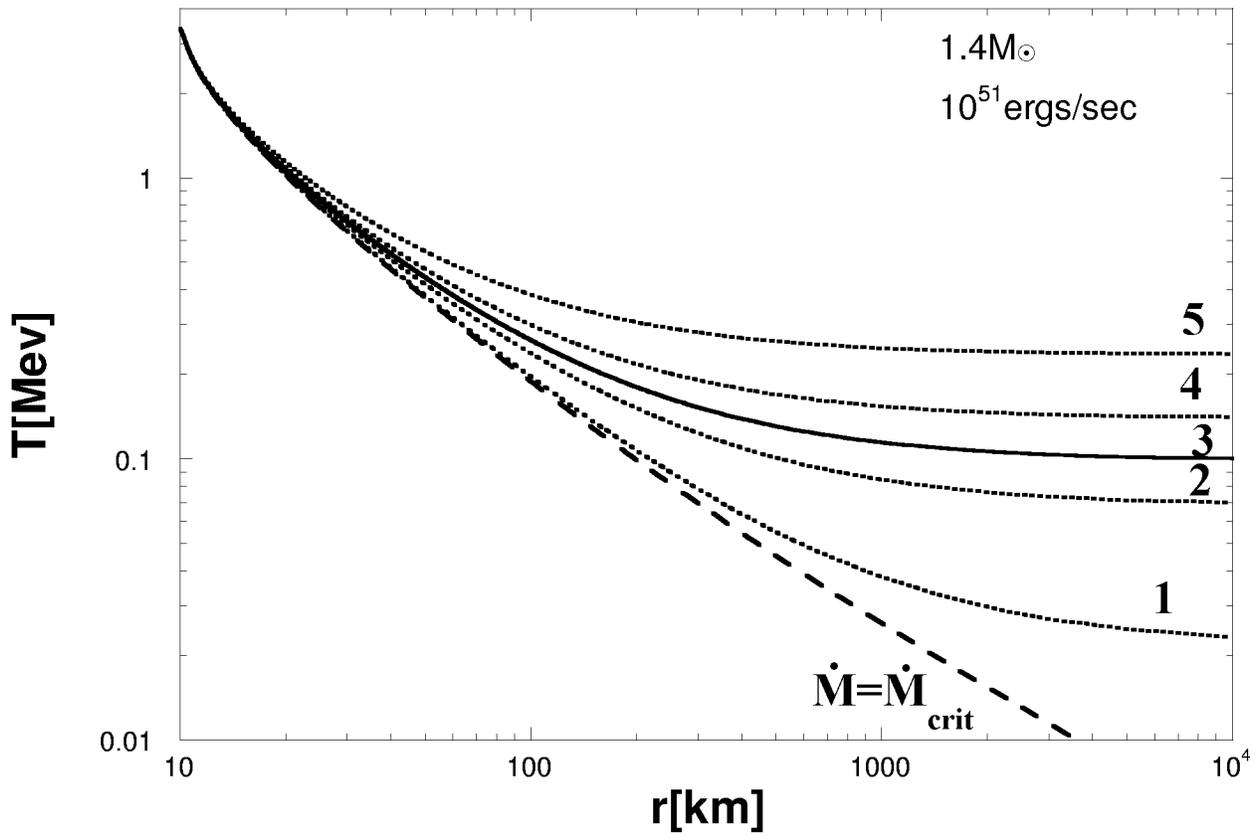
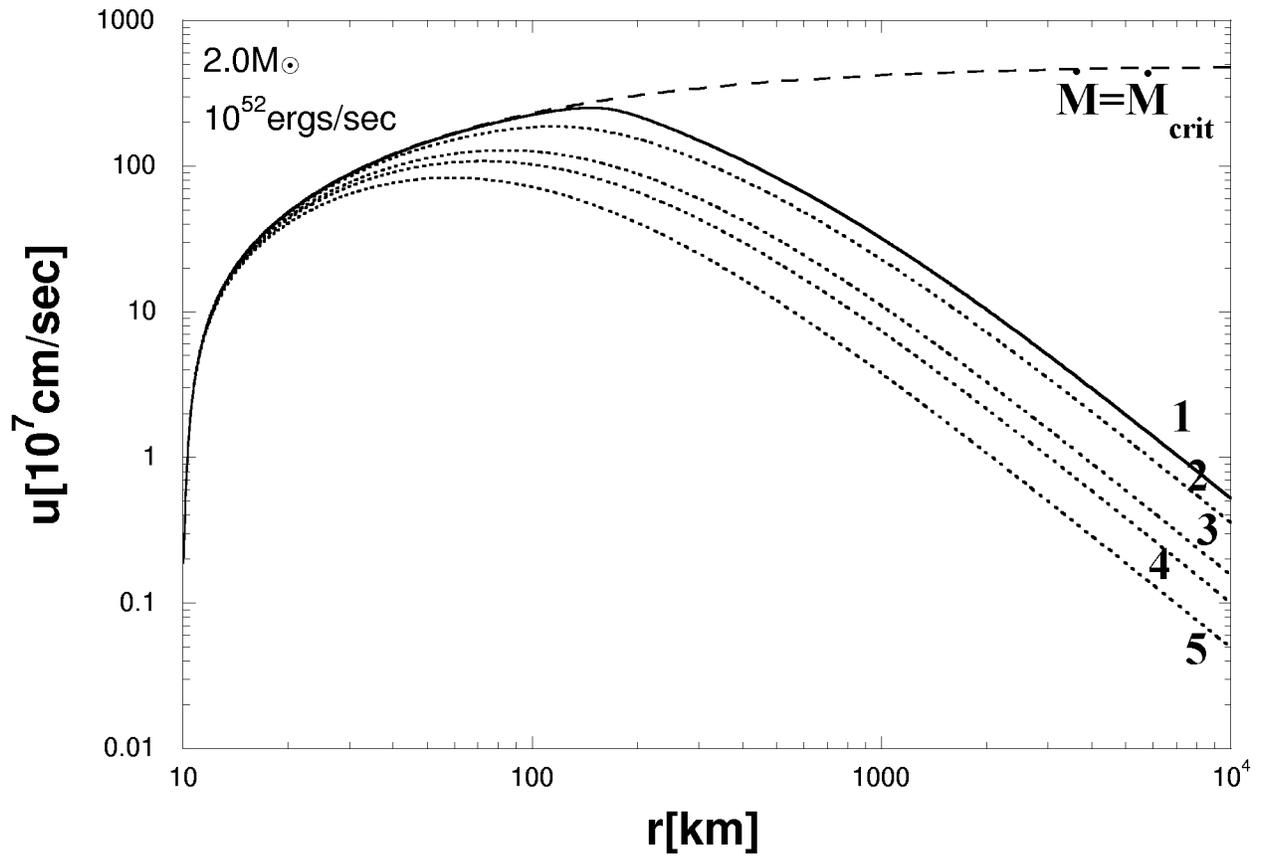
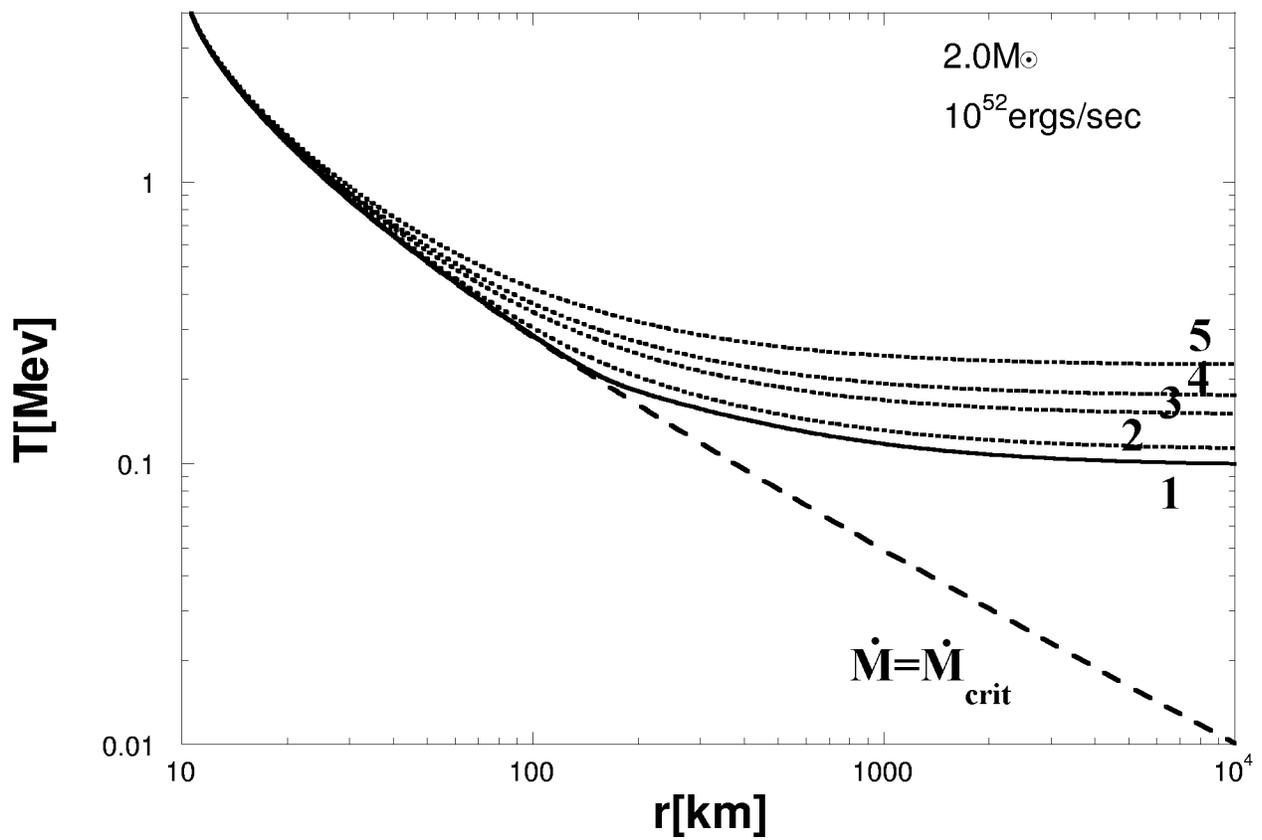
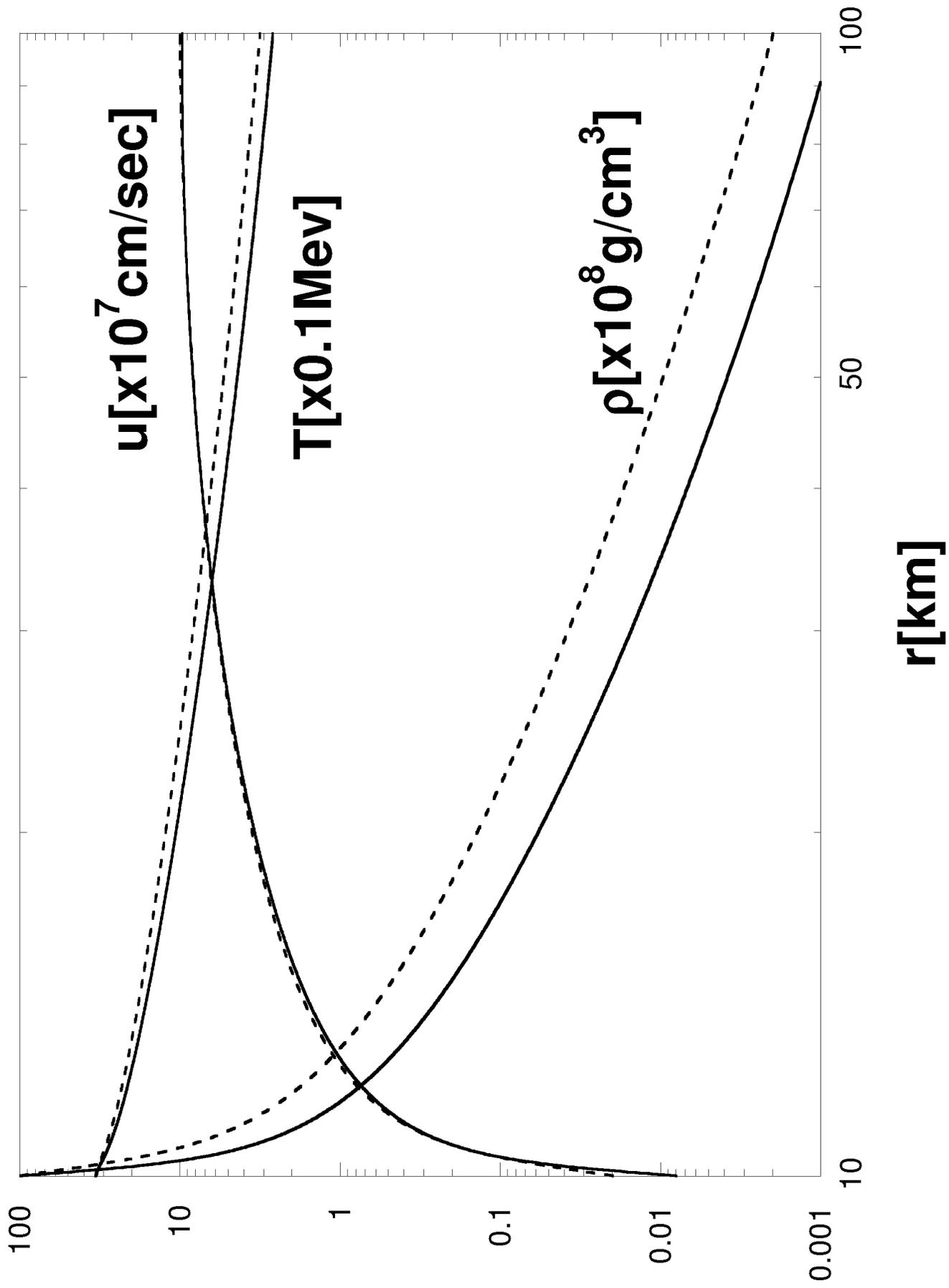
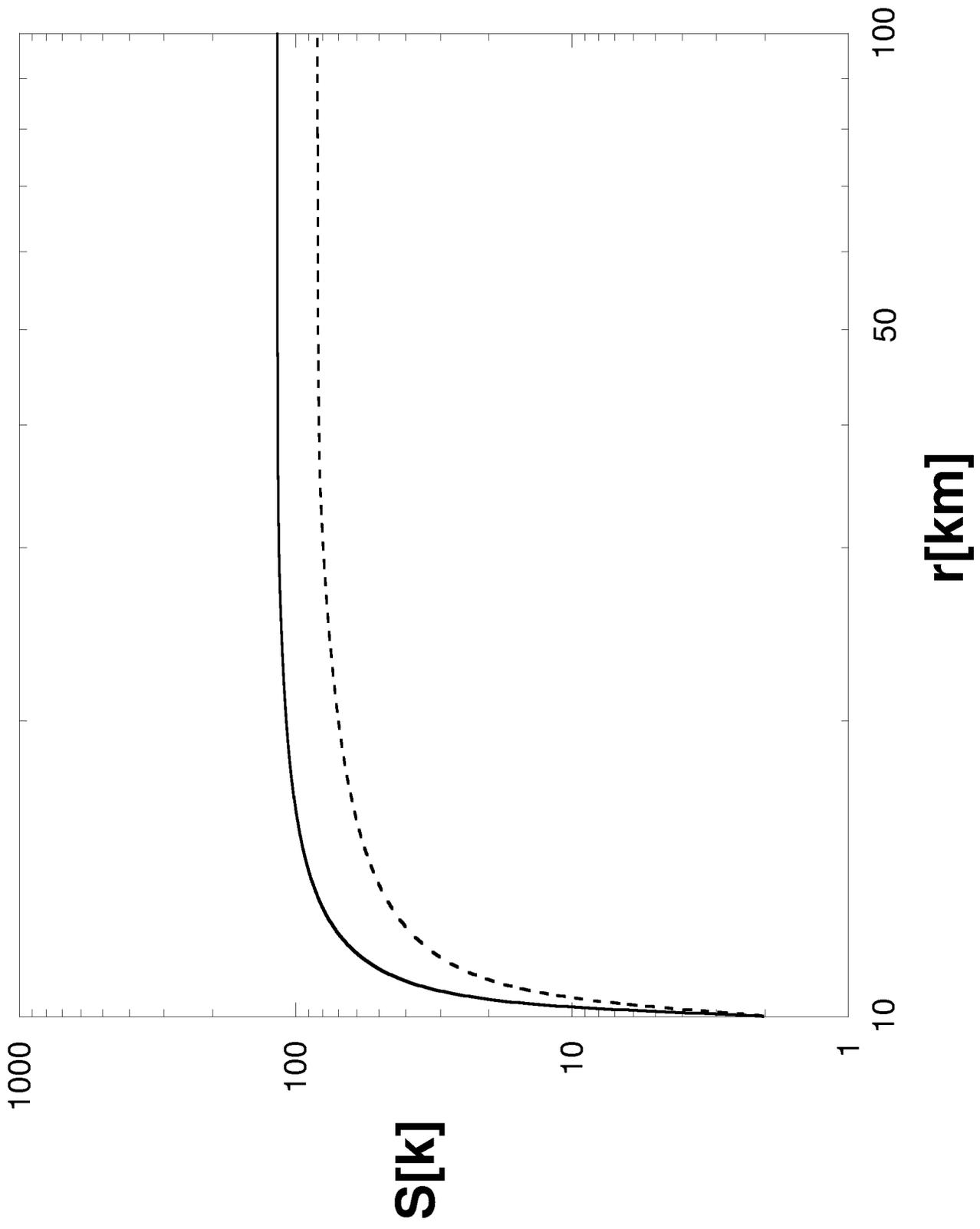
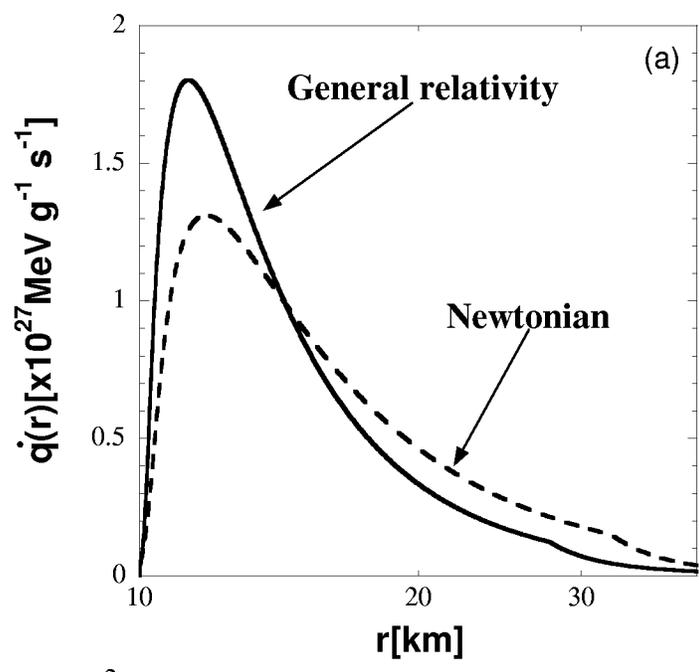
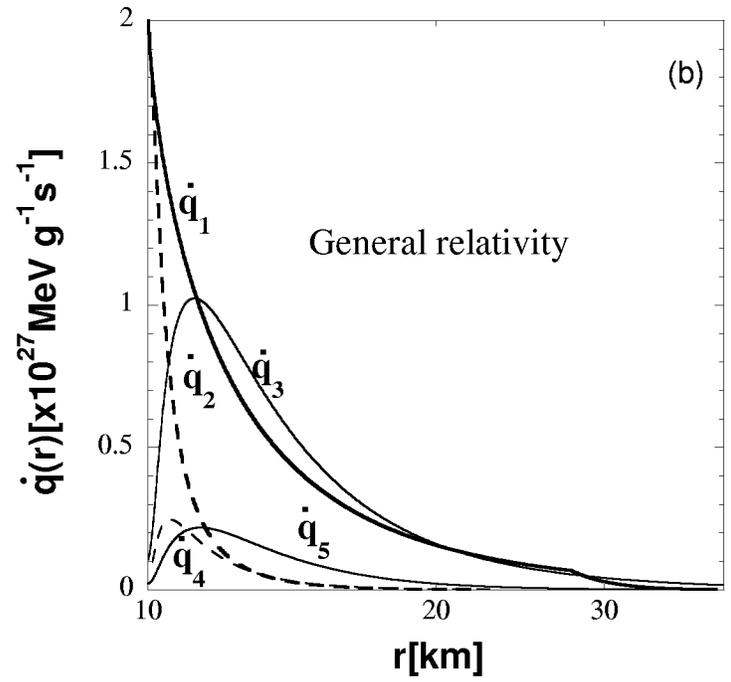
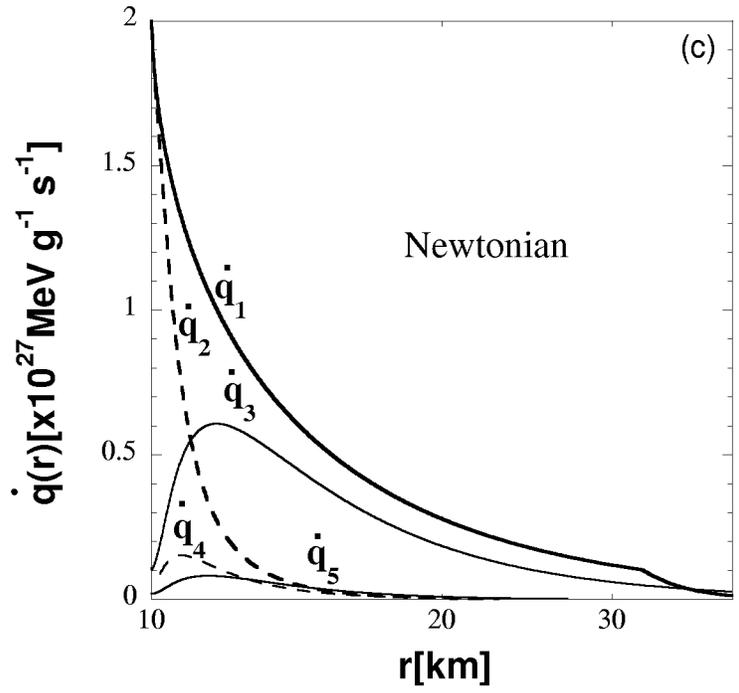
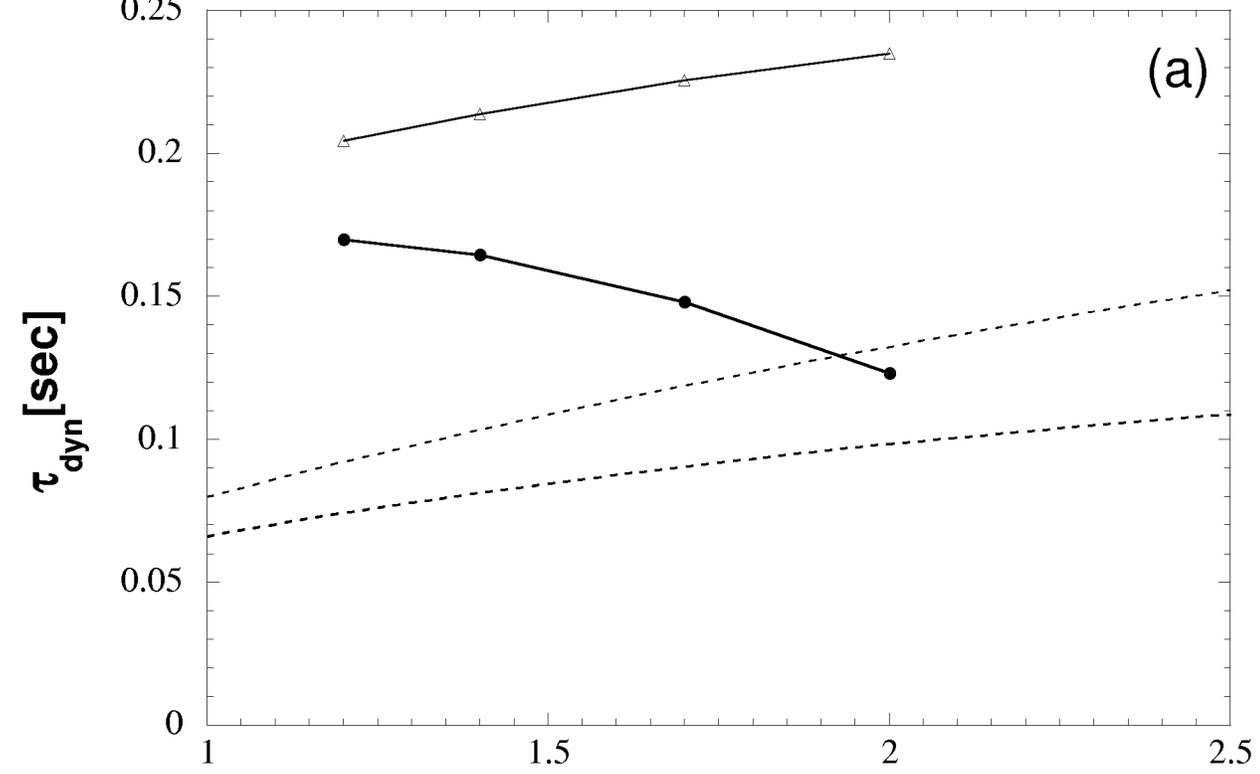
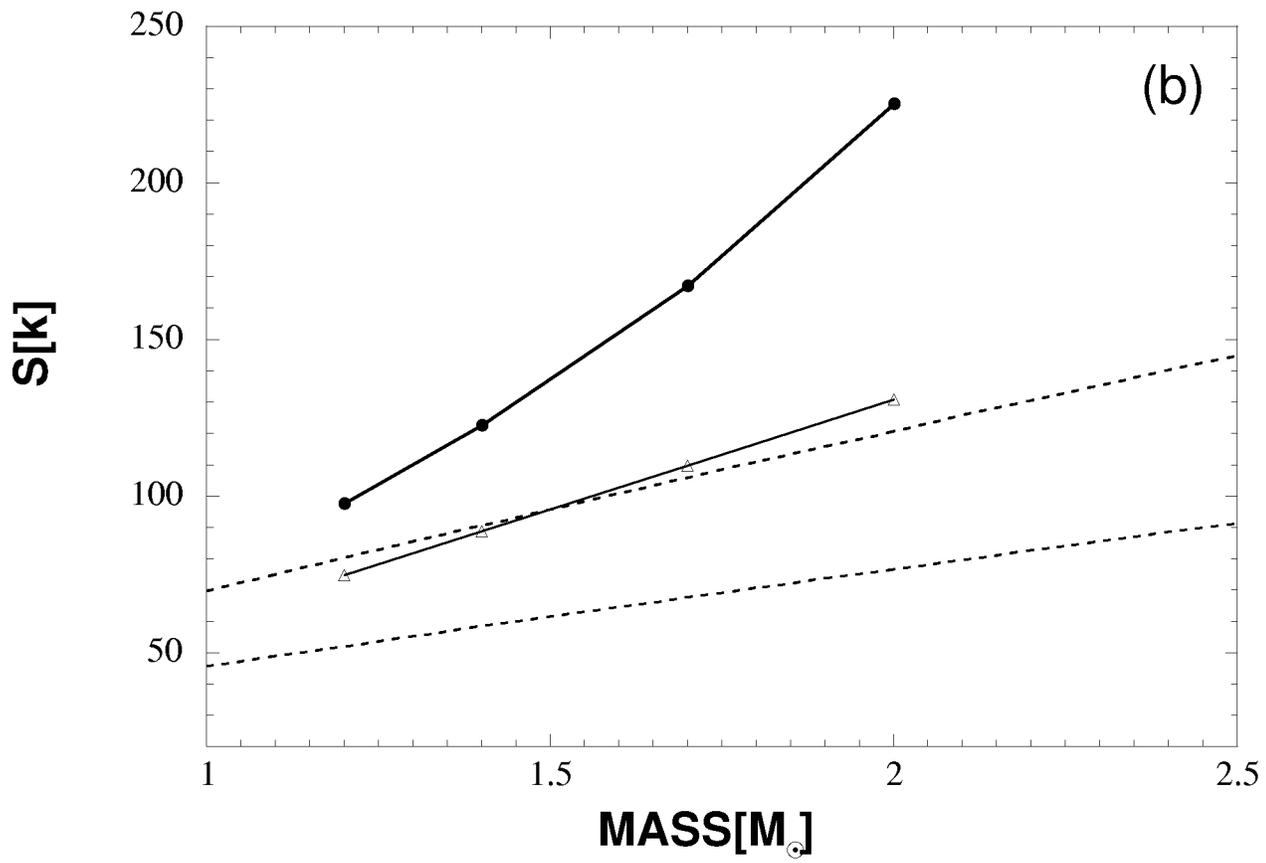
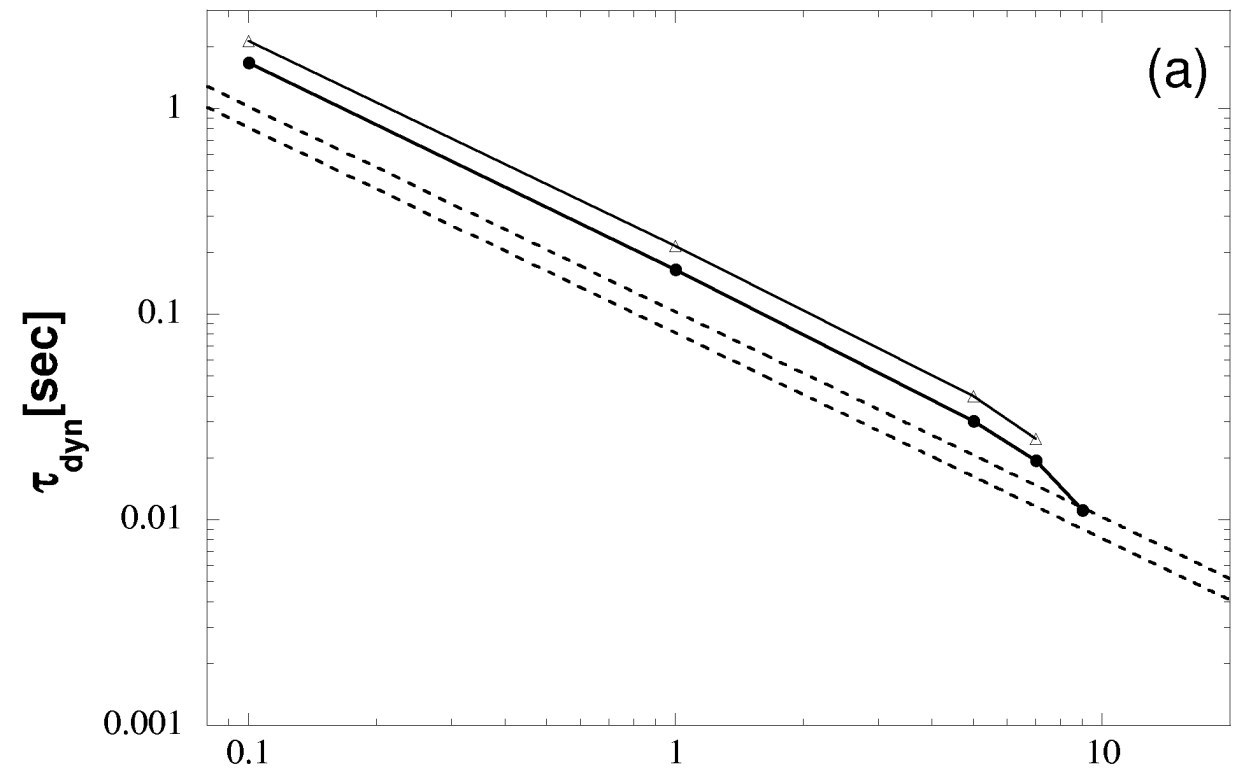
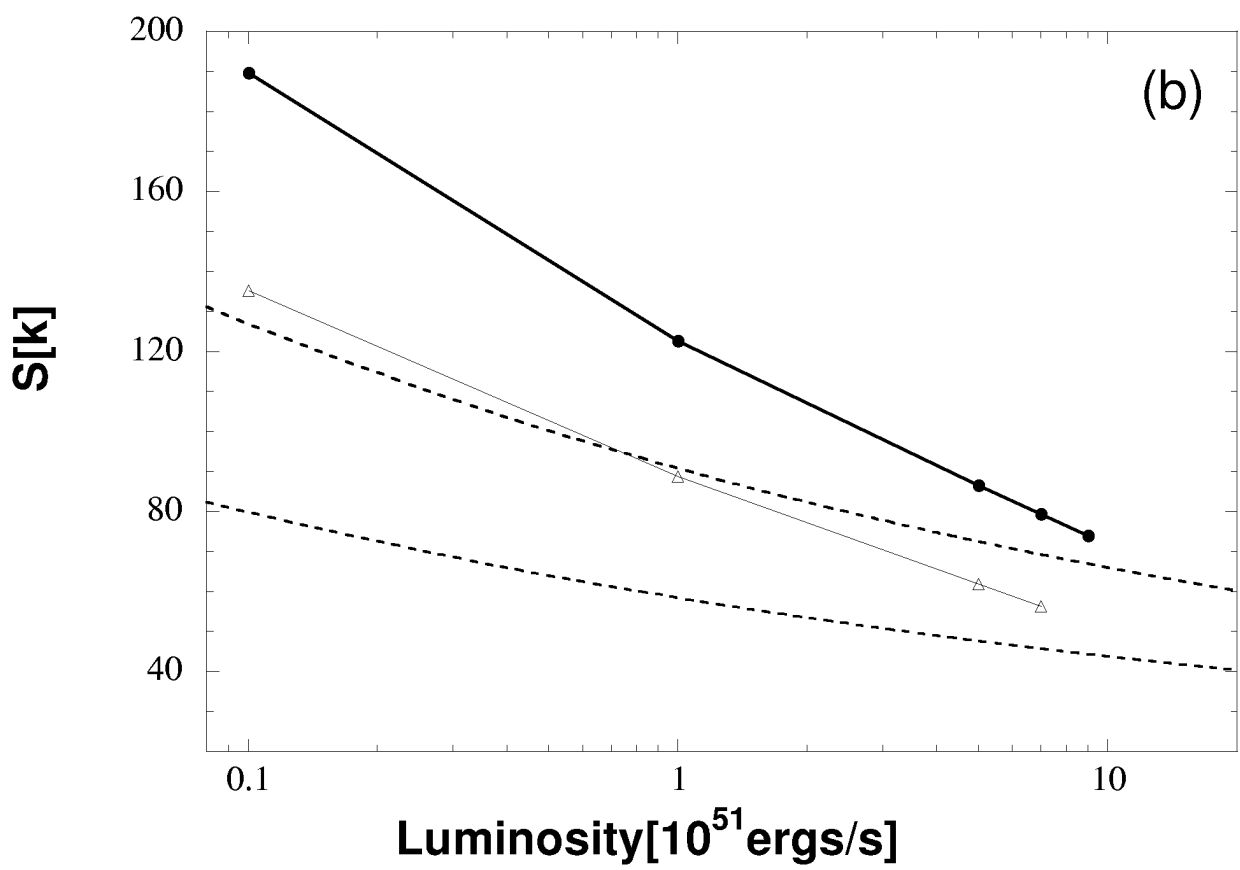
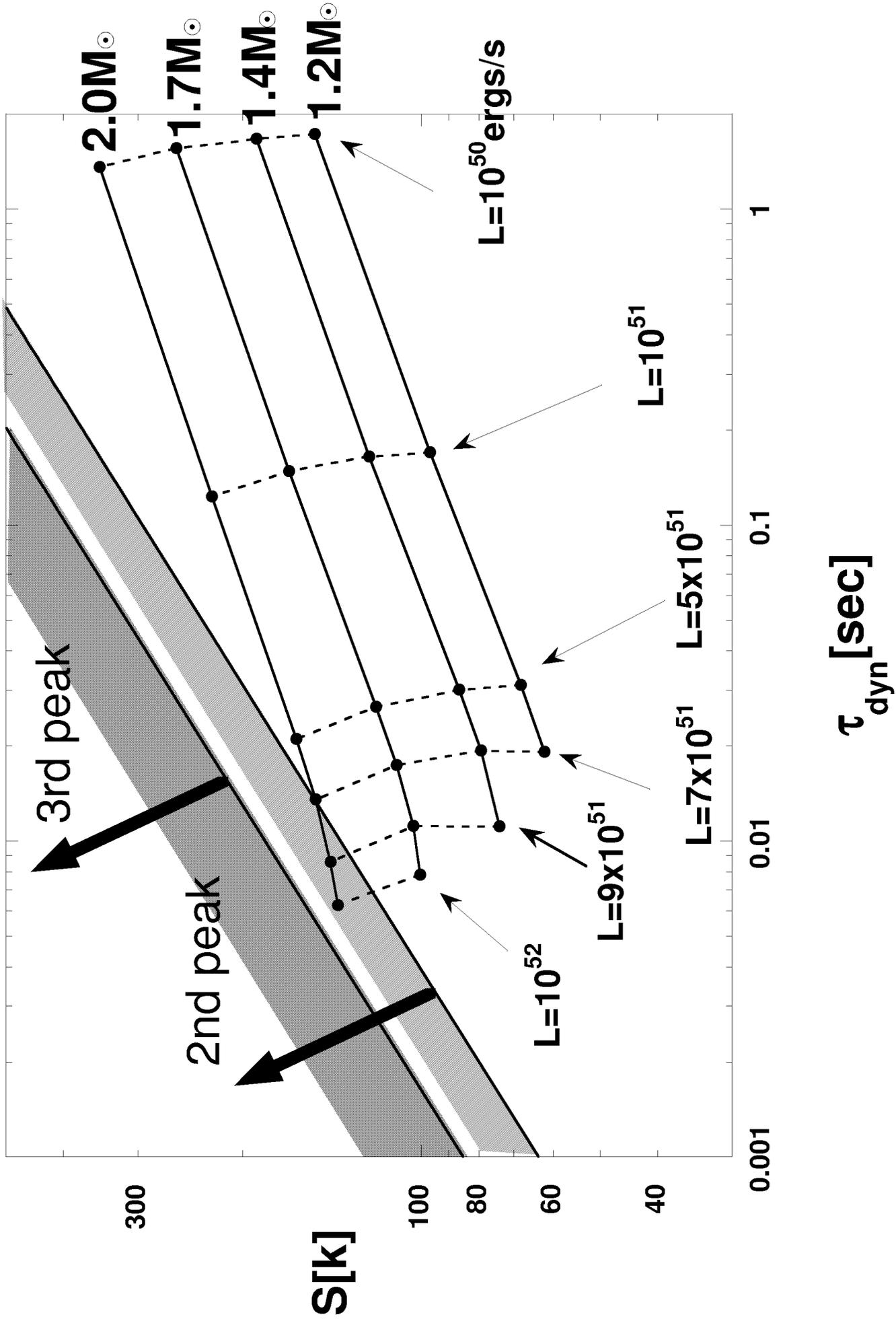
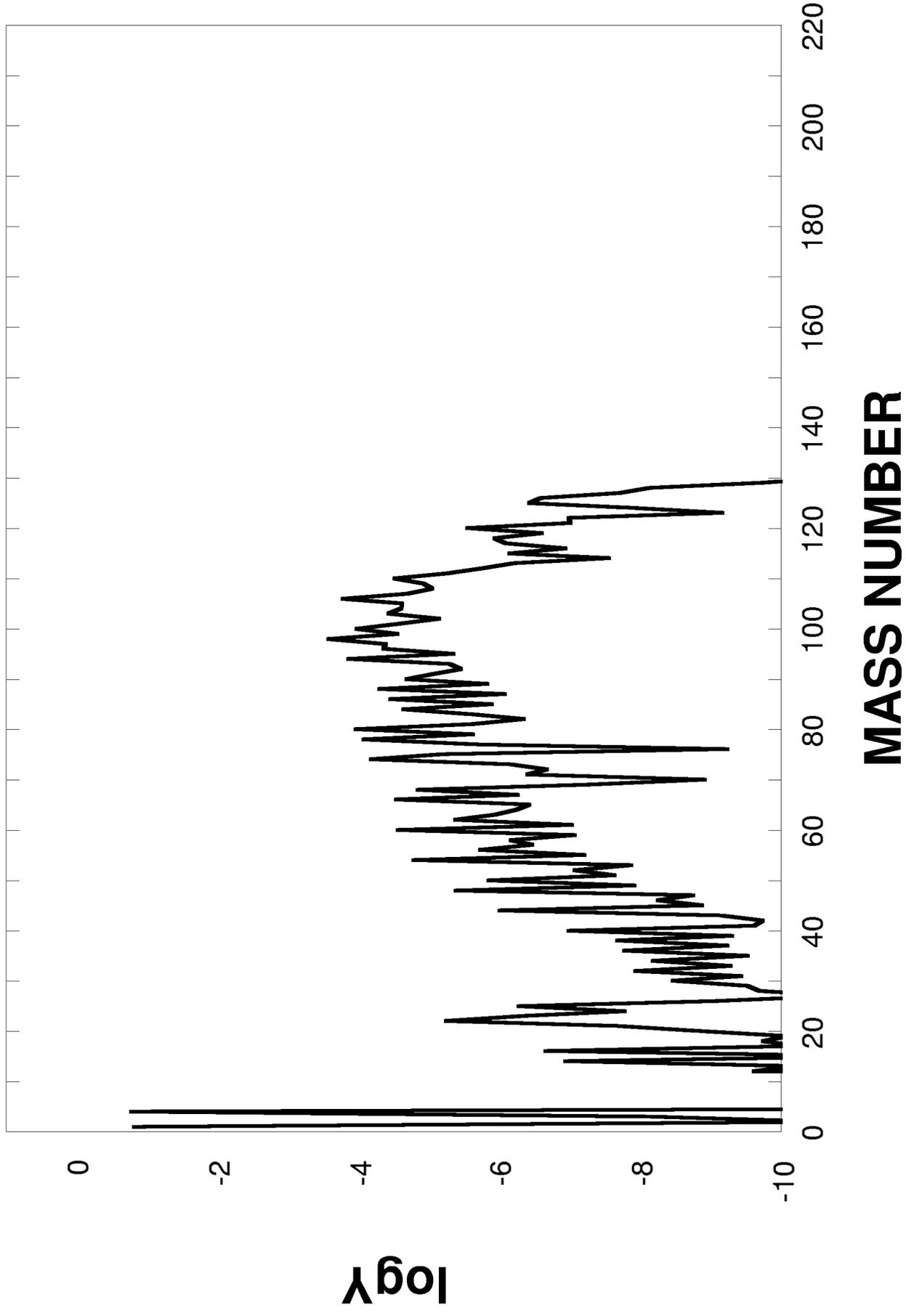
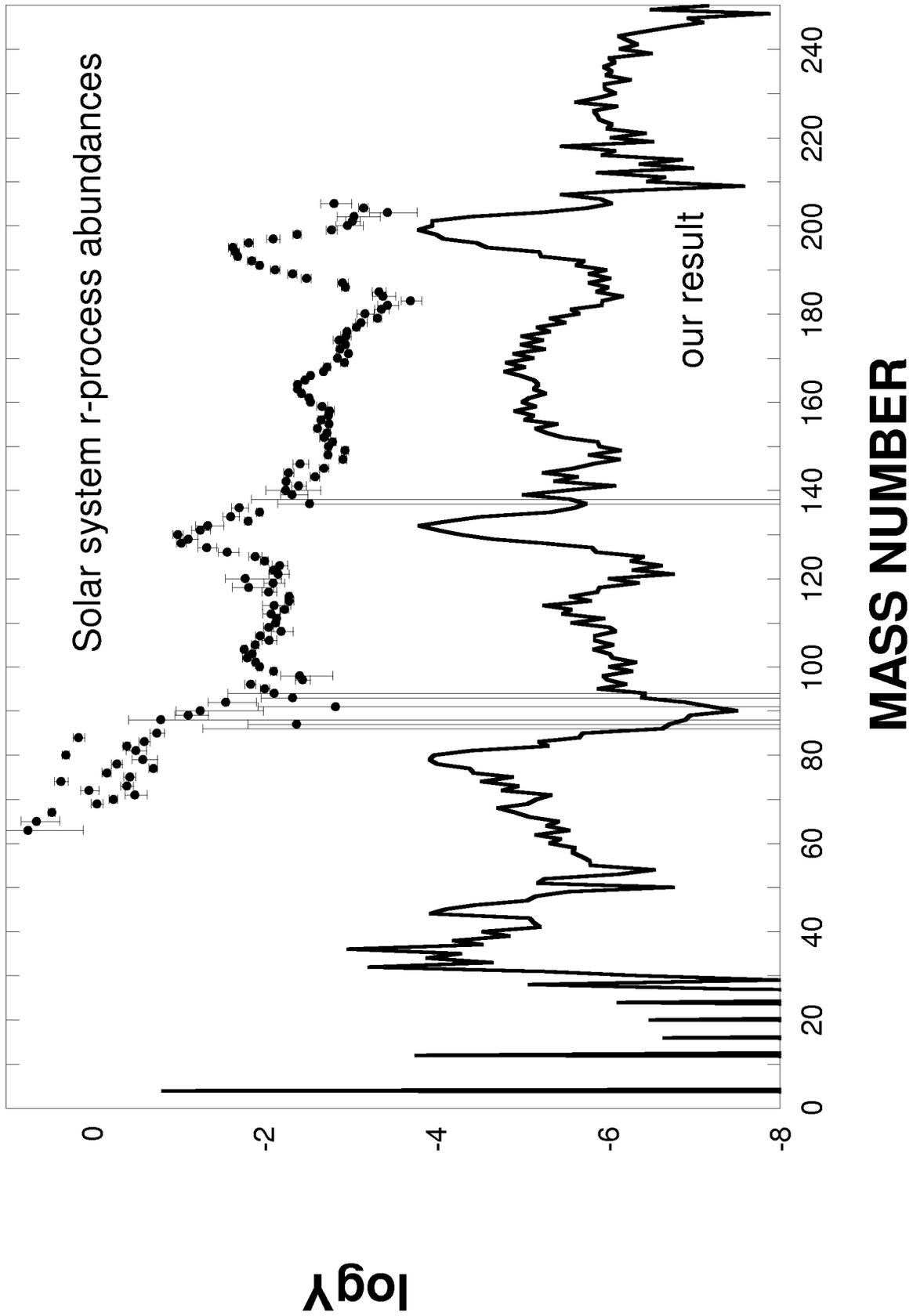
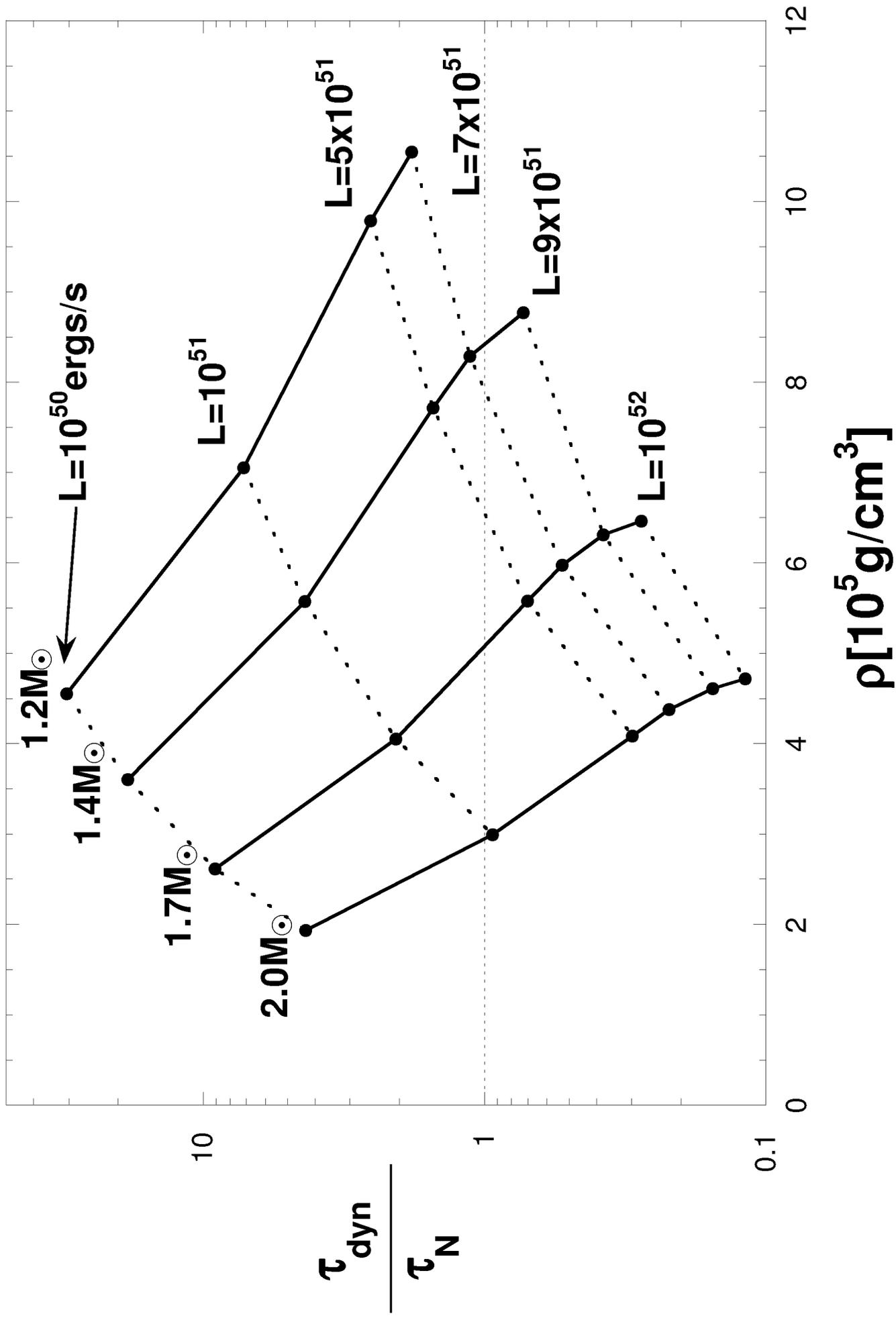
\end{document}